\documentclass[twocolumn,epsfig,superscriptaddress,prl]{revtex4-2}

\usepackage{graphicx}
\usepackage{amssymb}
\usepackage{amsmath}
\usepackage{hyperref}
\usepackage{changes}
\usepackage{comment}
\usepackage{mathtools,nccmath}
\usepackage{ragged2e} 
\usepackage{stackengine}
\usepackage{placeins}
\usepackage[percent]{overpic}
\usepackage{pgfplots}
\usepackage{tabularx}
\usepackage[caption=false]{subfig} 
\DeclareCaptionJustification{justified}{\justifying}

\usepackage{pgfplots}
\pgfplotsset{compat=1.8}
\DeclareMathAlphabet{\mathpzc}{OT1}{pzc}{m}{it}

\begin{document}
\title{AC Josephson Signatures of the Superconducting Higgs Mode}	

\author{Aritra Lahiri}
\email[]{aritra.lahiri@uni-wuerzburg.de}
\affiliation{Institute for Theoretical Physics and Astrophysics,
University of W\"urzburg, D-97074 W\"urzburg, Germany}
\author{Sang-Jun Choi}
\affiliation{Department of Physics Education, Kongju National University, Gongju 32588, Republic of Korea}
\author{Björn Trauzettel}
\affiliation{Institute for Theoretical Physics and Astrophysics, University of W\"urzburg, D-97074 W\"urzburg, Germany}
\affiliation{W\"urzburg-Dresden Cluster of Excellence ct.qmat, Germany}
\date{\today}

\begin{abstract}
{The Higgs mode in superconductors corresponds to oscillations of the amplitude of the order parameter. While its detection typically entails resonant optical excitation, we present a purely transport-based setup wherein it is excited in a voltage biased Josephson junction. Demonstrating the importance of order parameter dynamics, the interplay of Higgs resonance and Josephson physics enhances the second harmonic Josephson current oscillating at twice the usual Josephson frequency in transparent junctions featuring single-band s-wave superconductors. If the leads have unequal equilibrium superconducting gaps, this second harmonic component may even eclipse its first harmonic counterpart, thus furnishing a unique hallmark of the Higgs oscillations.}
\end{abstract}
\maketitle
The non-equilibrium dynamics of superconductors (SCs), resulting from an interplay of collective modes and quasiparticles, have been studied extensively~\cite{Chang1978,Kopnin2001,Tinkham1979}. Recent experimental developments have ushered in several insights, especially about the superconducting Higgs mode~\cite{Higgs1964,Littlewood1982,Varma2002,Pekker2015,Shimano2020,Matsunaga2012,Matsunaga2013,Matsunaga2014,Katsumi2018,Matsunaga2017}. The spontaneous $U(1)$ symmetry breaking in SCs results in a complex order parameter (OP) $\Delta(t) = |\Delta(t)|\exp(i\theta(t))$ with two collective modes: Higgs mode, corresponding to the amplitude $|\Delta(t)|$, and Goldstone mode, tied to the phase $\theta(t)$. While the Goldstone mode is pushed up to the plasma frequency~\cite{Anderson1958}, the Higgs mode has energy $\omega_H=2\Delta_0$~\cite{Varma2002,Pekker2015}, where $\Delta_0$ is the equilibrium gap amplitude. Being a scalar mode with no charge~\cite{Pekker2015,Schwarzthesis}, it has no linear coupling to the electromagnetic field. As such, its detection requires either strong lasers to harness non-linear electromagnetic effects made possible only by recent advances in THz technology~\cite{Matsunaga2012,Matsunaga2013,Matsunaga2014,Matsunaga2017,Katsumi2018,Sherman2015,Chu2020,Vaswani2021}, or coupling with coexisting electronic orders such as charge-density waves~\cite{Sooryakumar1980,Measson2014,Cea2014,Grasset2018}. Departing from this paradigm of purely optical techniques, there are limited transport-based proposals~\cite{Silaev2020,Tang2020,Vallet2023,Plastovets2023,Heckschen2022,Lee2023,Vallet2024}.

\begin{figure}\includegraphics[width=2.43in]{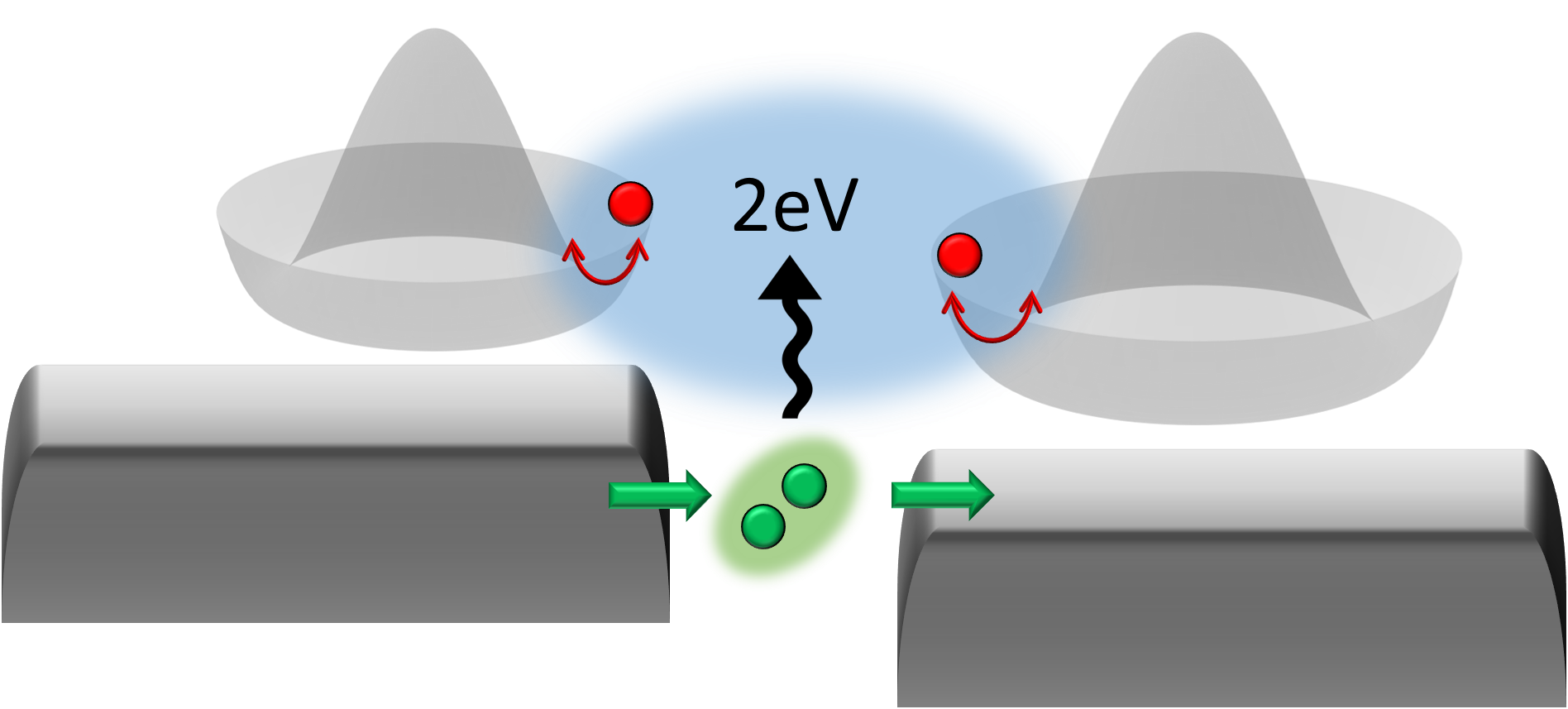}
\caption{Schematic of the Higgs mode, corresponding to radial oscillations of the OP (red balls) in the free-energy landscape (mexican hat). It is excited by radiating tunneling Cooper pairs (green balls) in a voltage-biased Josephson junction. We show the leads of the junction, which have lengths comparable to the superconducting coherence length to allow the Higgs oscillations to fully develop. The leads terminate in macroscopic superconducting reservoirs (not shown).}
\label{Fig1} 
\end{figure}

In this Letter, we investigate the Higgs mode in transparent voltage-biased Josephson junctions without external irradiation. The Josephson effect~\cite{Josephson1962} epitomises coherent phase dynamics, with the interference of SC condensates at the junction generating a supercurrent. For a constant voltage $V$, in the absence of amplitude oscillations, a supercurrent $\sim\sin(\omega_Jt)$ is generated solely by the condensate phase difference, where $\omega_J=2eV$ ($\hbar=1$) is the Josephson frequency. We find that the Josephson coupling leads to a dynamical inverse proximity effect, inducing OP amplitude oscillations by linking it to the energy emitted by tunneling Cooper pairs~\cite{Hofheinz2011} via the phase difference~\cite{Pekker2015,Klapwijk1982}. This obviates the need for external irradiation (see Fig. \ref{Fig1}). Microscopically, the pairs coherently dissociate into Bogoliubov quasiparticles over the energy gap $\Delta_0$ and recombine, requiring energy $\geq 2\Delta_0$. Consequently, a resonance emerges at $\omega_J=\omega_H$. While the possibility of probing the collective modes of a fluctuating superconductor slightly above its critical temperature using \emph{DC} pair tunneling is established~\cite{Ferrell1969,Scalapino1970}, recently this was extended to zero temperature in a Josephson setup~\cite{Lee2023}. However, the Higgs mode was obfuscated by multiple Andreev reflections appearing at the same voltage in the \emph{DC} current-voltage characteristics. We investigate instead the \emph{AC} Josephson response, finding that the interplay of varying condensate phases and amplitudes generates a strong current at frequency $2\omega_J$. Remarkably, it can supercede the typical $\omega_J$ current when the leads' equilibrium gaps differ significantly, constituting a clear indication of Higgs oscillations in time-reversal symmetric Josephson junctions featuring single-band s-wave SCs~\cite{trsnote,Sellier2004,Sickinger2012,Goldobin2007,Frolov2006,Li2019,Yokoyama2014,Tanaka1997,Ng2009,Asano2003,Trimble2021,Can2021,Tummuru2022,Sun2024}.

\textit{Phenomenology.--} For an intuitive picture, we start with the phenomenological zero-temperature effective field theory of the superconducting OP ~\cite{Varma2002,Pekker2015,Schwarzthesis,Kleinert2018,vanOtterlo1999,Tsuji2015,Hansson2004,Schwarz2021,Haenel2021,Puviani2020,Altlandbook}. On expanding about the equilibrium OP values, $\Delta_{L/R}=\Delta_{0,L/R}+\delta\Delta_{L/R}$, the Gaussian Lagrangian density is given by $\mathcal{L}=\sum_{j=L,R}\mathcal{L}_j+\mathcal{L}_J$, where,
\begin{equation}
\begin{split}
\mathcal{L}_j=&(\partial_t\delta\Delta_j)^2-\vartheta^2(\partial_x\delta\Delta_j)^2-\omega_{H,j}^2\delta\Delta_j^2,\\
\mathcal{L}_J=&- 2J\Delta_L\Delta_R\cos(\omega_J t).
\end{split} \label{effFT}
\end{equation}
Here, $\Delta_{L/R}\in \Re$ are the OPs in the left/right leads, $\vartheta\sim$ the superfluid density, and $J\sim\mathcal{T}^2$ where $\mathcal{T}$ parametrises the coupling across the junction. We use a gauge where the gaps are real, and the Josephson phase accounts for the voltage-dependent condensate phases. Particle-hole symmetry~\cite{Varma2002,Pekker2015,phsnote,Phan2023} dictates the dynamical term $(\partial_t\delta \Delta)^2$. The Higgs mass $\omega_{H,L/R}=2\Delta_{0,L/R}$~\cite{Varma2002,Pekker2015,Kleinert2018} results from microscopic calculations. The non-equilibrium OP corrections $\delta\Delta_{L/R}$ satisfy
\begin{align}
\iint_{t',x'} \begin{bmatrix}
\chi_{\Delta\Delta,L}^{-1} &  s'_{\phi,L}\\
 s'_{\phi,R} & \chi_{\Delta\Delta,R}^{-1}
\end{bmatrix}_{(t,t';x,x')} \begin{bmatrix} \delta \Delta_L\\ \delta \Delta_R\end{bmatrix}_{(t';x')} = \begin{bmatrix}s_{\phi,L} \\ s_{\phi,R}\end{bmatrix}_{(t;x)}\label{ddelgl}
\end{align}
with $\chi^{-1}_{\Delta\Delta,L/R}=(\partial_t^2+\omega_{H,L/R}^2-\vartheta^2\partial_x^2)\delta(t-t')\delta(x-x')$, $s_{\phi,L/R}=-J\Delta_{0,R/L}\cos(\omega_J t)\delta(x)$, and $s'_{\phi,L/R}=J\cos(\omega_J t)\delta(t-t')\delta(x)$. Eq. \eqref{ddelgl} resembles driven coupled oscillators. The source term $s_\phi$ encapsulates the energy emitted by tunneling pairs. Thus, it oscillates at $\omega_J$. The cross terms $s_\phi'$ describe the coupling between the oscillating OPs, which are driven by the junction field with $s'_\phi$ oscillating at $\omega_J$. At leading order in $J$, neglecting the cross-coupling, $\delta \Delta_{L/R}^{(J)}(t) = \int_{-\infty}^t dt'\chi^r_{\Delta\Delta,L/R}(t,t')s_{\phi,L/R}(t')$. An explicit calculation for $\omega_J<\omega_{H,L/R}$ reveals $\delta\Delta_{L/R}(t,x)= (-J\Delta_{0,R/L}/(2\vartheta|\omega_H^2-\omega_J^2|^{0.5}))\cos(\omega_Jt)e^{-|\omega_H^2-\omega_J^2|^{0.5}|x|/\vartheta} $, deriving its resonant enhancement from the Higgs pole in $\chi_{\Delta\Delta,L/R}(\omega_J,q)=1/(-\omega_j^2+\omega_{H,L/R}^2+\vartheta^2q^2)$. A similar calculation for $\omega_J\geq\omega_{H,L/R}$ depends on the microscopic damping, arising from the decay of the Higgs mode into quasiparticles. Still, we find that $\delta\Delta$ oscillates at $\omega_J$, with its spatial decay into the bulk governed by the damping. Within this simple model, the current may be approximated as $I\sim J\Delta_{L}(t)\Delta_{R}(t)\sin(\omega_J t)$, which is sensitive to the dynamics of the OP amplitudes. 

We discover that, for $\Delta_{0,L}\neq\Delta_{0,R}$, the Higgs oscillations in the OP manifest as a Josephson current with an enhanced $2\omega_J$ component, which may even dominate the usual $\omega_J$ component. Considering $\Delta_{0,L}<\Delta_{0,R}$ without loss of generality, as the voltage is increased and $\omega_J$ approaches $\omega_{H,L}$, $\delta\Delta_{L}$ is Higgs-enhanced while the non-resonant $\delta\Delta_{R}$ is small. In this regime, where $\Delta_L=\Delta_{0,L}+\delta\Delta_L$ and $\Delta_R\approx\Delta_{0,R}$, we find a $2\omega_J$ component in the current $I_{2\omega_J}$ bearing the amplitude $I_{2\omega_J}=I_{\omega_J}f_H$, where $I_{\omega_J}=J\Delta_{0,L}\Delta_{0,R}$ is the amplitude of the usual $\omega_J$ component, and $ f_H\sim-(J/(4\vartheta |\omega_{H,L}^2-\omega_J^2|^{0.5}))(\Delta_{0,R}/\Delta_{0,L})$ reflects the Higgs enhancement. At resonance, $I_{2\omega_J}$ is bounded only by the Higgs lifetime. 

For equal equilibrium gaps $\Delta_{0,L}=\Delta_{0,R}=\Delta_0$, instead, both OPs reveal Higgs oscillations with $\delta\Delta_{L}=\delta\Delta_{R}$. A similar analysis shows in this case that both $I_{2\omega_J}$ and $I_{\omega_J}$ get Higgs enhanced, precluding an outright dominance of the $2\omega_J$ current. 

Finally, despite the simplicity and intuitiveness of this phenomenological approach~\cite{Pekker2015}, it is an equilibrium formulation, applicable only approximately for \emph{tunnel} junctions for sub-gap voltages, in the absence of dissipative quasiparticle tunneling~\cite{Ambegaokar1982,Eckern1984,Schon1990}. In the present scenario, where the OP oscillates at the same time scale as $\phi$ and decays into the quasiparticle continuum~\cite{Littlewood1982}, it is imperative to account for the proper retarded dynamics with a microscpic approach~\cite{Larkin1967,Ambegaokar1982,Werthamer1966,Lahiri2023}. Nevertheless, a microscopic perturbative approach yields an equation similar to Eq.~\eqref{ddelgl} (see the Supplemental Material (SM)~\cite{sm}).

\begin{figure*}[!htb]
\includegraphics[width=\textwidth]{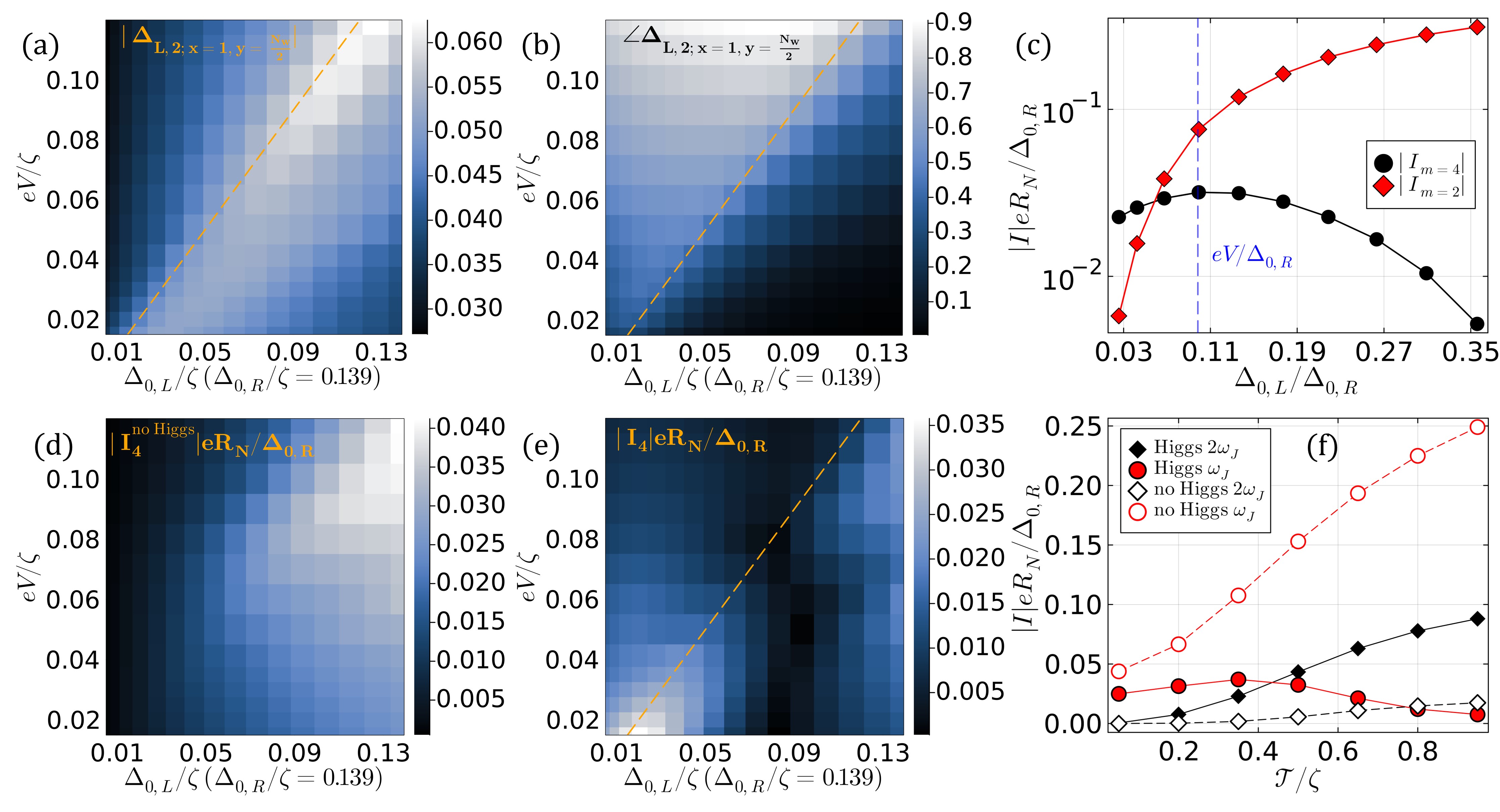}
\caption{Numerical results, considering a two-dimensional junction with $N_{L}=36$, $N_{R}=6$, $N_w=12$, $\mu=0$, $\mathcal{T}=0.4\zeta$ (normal state transparency$\ \approx 0.48$~\cite{Cuevas1996}), $\Gamma=0.02\zeta$, and $\zeta=5$ (bandwidth$=40$). We define the SC coherence length of the right lead (remains unchanged) $\xi_{sc,R}=4\zeta/(\pi\Delta_{0,R})$. We show the OP at the $y=6^{\text{th}}$ site along the transverse direction in (a-b). The OP shows minor transverse variations~\cite{sm}. (a) $|\Delta_{2}|$ (frequency $\omega_J=2(eV)$) at the first site on the left SC immediately neighbouring the junction, showing the Higgs resonance at $eV=\Delta_{0,L}$ (orange dashed line). (b) Same as (a), but we show the argument $\angle\Delta_{2}$, such that the OP modulation is $2|\Delta_2|\cos(\omega_J t-\angle\Delta_{2})$, revealing a jump across the resonance (orange dashed line). In (c-f), we define $R_N$ as the numerically obtained normal state resistance. (c) The current components $I_4$ (frequency $2\omega_J=4(eV)$) and $I_2$ (frequency $\omega_J=2(eV)$), at $eV/\zeta=0.02$, on a logarithmic scale. With decreasing $\Delta_{0,L}/\Delta_{0,R}$, $I_4$ strengthens and peaks near the Higgs resonance ($eV/\Delta_{0,R}=\Delta_{0,L}/\Delta_{0,R}$ blue dashed line), eventually dominating $I_2$. (d) Amplitude of $I_4$ with varying voltage and $\Delta_{0,L}$ in the absence of Higgs oscillations, forcing $\delta\Delta=0$. (e) Same as (d), but with Higgs oscillations included. $I_4$ exhibits a peak at the Higgs resonance (orange dashed line) for highly asymmetric junctions (bottom left corner of the plot)---a feature absent in panel (d). The slight deviation from $eV=\Delta_{0,L}$ likely stems from the enhancement of the equilibrium gap near the junction from its bulk value $\Delta_{0,L}$ via the inverse proximity effect~\cite{sm}. (f) $I_4$ (black diamonds) and $I_2$ (red circles) are shown as functions of $\mathcal{T}$, both with (solid markers) and without (empty markers and dashed lines) Higgs oscillations, for $\Delta_{0,L}/\Delta_{0,R} \approx 0.07$ and $eV/\zeta = 0.02 \approx \Delta_{0,L}/\zeta$. Notably, $I_4$ exceeds $I_2$ at large $\mathcal{T}$ only in the presence of the Higgs enhancement. }
\label{Fig2} 
\end{figure*}

\textit{Model.--} We consider $s$-wave BCS SC Hamiltonians~\cite{AGD1975,deGennes1966,Stefanucci2010,Yeyati1995}. The device consists of two SC leads of length $L$ satisfying $\lambda_F\ll L\sim \xi_{\text{sc}}$ where $\lambda_F$ is the Fermi wavelength and $\xi_{\text{sc}}$ is the coherence length. This is because the Higgs modulations occur over a length-scale $\sim \xi_{\text{sc}}$. Since we do not explicitly model electromagnetic screening, our results are, in principle, valid for junctions with widths comparable to the electromagnetic penetration depth. However, the time-dependent phenomena obtained in this work should remain qualitatively unaffected even for wider junctions. The outer ends of the leads are connected to macroscopic SC reservoirs. For computational simplicity, here we focus on a two-dimensional planar junction with specular tunneling. A similar three-dimensional system, assuming a uniform OP along the transverse directions, is analysed in the SM~\cite{sm}. The results confirm that our conclusions remain unchanged. The left (L) and right (R) leads contain $N_{L}$ and $N_{R}$ sites, respectively, along the longitudinal direction, and $N_w$ sites along the transverse direction. Both leads are connected to infinite superconducting reservoirs. The mean-field Hamiltonian is $H=H_L+H_R+H_{\mathcal{T}}$, where,
\begin{align}
H_{L/R}= &\hspace{-1.0em}{\textstyle\sum\limits_{\substack{j_{x,y}\in L/R\\ \sigma}}}\big[ \big(-\zeta c_{j_x+1,j_y,\sigma}^\dagger c_{j_x,j_y,\sigma}-\zeta c_{j_x,j_y+1,\sigma}^\dagger c_{j_x,j_y,\sigma}\nonumber\\
&\hspace{1.162cm}+\Delta_{j_x,j_y}(t)c^\dagger_{j_x,j_y,\sigma}c^\dagger_{j_x,j_y,\sigma'} +h.c.\big],\label{Ham}\\
H_{\mathcal{T}}=&{\textstyle\sum\limits_{j_y,\sigma}} \big[-\mathcal{T}e^{i\frac{\phi(t)}{2}}c_{L1,j_y,\sigma}^\dagger c_{R1,j_y,\sigma} + h.c.\big].\label{HamT}
\end{align}
Here, $j_{x/y}$ indexes sites along the longitudinal/transverse direction, $\zeta$ is the hopping amplitude (bandwidth $=8\zeta$), $\mathcal{T}$ is the junction coupling~\cite{Cuevas2002}, and $\phi(t)=\int_{-\infty}^t d\tau 2eV(\tau)$ (second Josephson relation). We use a gauge which shifts the voltage into the junction coupling~\cite{Cuevas1996}. We assume that the phase of the OP was spatially and temporally uniform inside the leads in the initial gauge, which results in a real OP ($\Delta(t)$) following the gauge transformation. Regarding the spatial uniformity, recalling that the bulk supercurrent admits the expression $J_s=(en_s/m)\nabla\phi$ where $n_s$ is the superfluid density, on ensuring current continuity, it can be shown that the variation of the phase over a distance $\xi_{\text{sc}}$ satisfies $\xi_{\text{sc}}\nabla\phi \sim \tau$, where $\tau=4(\mathcal{T}/\zeta)^2/(1+(\mathcal{T}/\zeta)^2)^2$ ($0<\tau<1$) is the junction transparency. As we show below, while the Higgs-enhanced Josepson current benefits from high junction transparency, it is enhanced by high equilibrium gap asymmetry $\Delta_{0,L}\neq\Delta_{0,R}$ as well. Thus, the requirement for high transparency $\tau$ can be relaxed by employing a larger superconducting gap asymmetry. As such, we can approximately neglect the spatial variations of the phase. Note that in the SC reservoirs the phase gradient vanishes as $\xi_{\text{sc}}\nabla\phi\sim (\mathcal{N}_J/\mathcal{N}_b)\tau\to 0$ where $\mathcal{N}_J$ and $\mathcal{N}_b$ denote the number of channels in the leads and the reservoirs, respectively~\cite{Beenakker1991a}. The temporal uniformity typically holds when electric fields are restricted to the barrier and penetrate into the leads over a distance comparable to the electromagnetic penetration depth, which is much smaller than $\xi_{\text{sc}}$ in type I SCs~\cite{Artemenko1997,Hsiang1980,Rieger1971,Kulikov2019}.

We use the Keldysh-Gorkov framework for a self-consistent solution to the OP~\cite{Collado2019,Kamenev2011,Kemoklidze1966,Volkov1974,Vadimov2019,Kuhn2024,Dzero2024,Kamenev2011}. We require the lesser Green's function which, in the Nambu space, bears the form $G^<_{\alpha\beta}(t,t')=-i\langle \psi_\alpha(t)\psi_\beta^\dagger(t')\rangle$, where $\psi_\alpha=[c_{\uparrow,\alpha}\ c^\dagger_{\downarrow,\alpha}]^T$ is the Nambu spinor, and $\alpha,\beta$ encapsulate spatio-temporal degrees of freedom. The non-equilibrium gap equation is given by,
\begin{equation}
\Delta_j(t)=igF^<_{j,j}(t,t),  \label{gapeq}
\end{equation}
where $g>0$ is the BCS coupling, and $F^<$ is the anomalous component of $G^<$.

Owing to the time-periodicity of Eq. \eqref{HamT} for DC voltage, we use the Floquet technique~\cite{Cuevas2002,Kuhn2024,Martinez2003,Stefanucci2008,Kalthoffthesis,SanJose2013,Bolech2005,Gavenskythesis,Lahiri2025}. The two-time Green's functions are expanded as
\begin{equation}
G(t,t')=\sum_{m,n} \int_0^\Omega \frac{d\omega}{2\pi} e^{i(\omega+n\Omega)t'-i(\omega+m\Omega)t}G_{mn}(\omega),
\end{equation}
where $\Omega=2\pi/T=eV$ is the fundamental Floquet frequency, $G$ is a matrix in the combined position and Nambu space, and $G_{mn}(\omega+l\Omega)=G_{(m+l)(n+l)}(\omega)$. The Dyson equations become~\cite{Jauho1994,Xu2019,Keldysh1964,Stefanuccibook2013,Gonzales2020,Gldecaynote,Collado2019},
\begin{equation}
\begin{split}
G^<_{mn}(\omega)&=G^r_{ml}(\omega)\Sigma^<_{ll'} (\omega)G^a_{l'n}(\omega),\\
G^{r/a}_{mn}(\omega)&=g^{r/a}_{mn}(\omega)+g^{r/a}_{ml}(\omega)\Sigma^{r/a}_{ll'}(\omega)G^{r/a}_{l'n}(\omega),
\end{split}\label{KDyson}
\end{equation}
where we have used the Einstein summation convention, $G^{r/a}$ is the retarded/advanced Green's function, $g^{r/a}_{mn}(\omega)=g^{r/a}(\omega+m\Omega)\delta_{mn}$ is defined in the absence of tunneling, and $\Sigma$ is the self-energy which we detail below. Following standard procedure~\cite{Samanta1998}, we partition the system into two leads connected at the junction, which we treat exactly, and two semi-infinite reservoirs attached to the far-ends of the leads. The self-energy consists of four terms: (i) tunneling self energy $\Sigma^{r/a}_{\mathcal{T},RL,m-n}=\Sigma^{r/a^{\scriptstyle \ast}}_{\mathcal{T},LR,n-m}=-[V_{m-n}(\tau_0+ \tau_3)/2 - V_{-m+n}^*(\tau_0- \tau_3)/2]$, where $V_{m-n}=(1/T)\int (d\omega/(2\pi))e^{i(m-n)\Omega t} \mathcal{T}e^{-i\phi(t)/2}$, with $\tau_\mu$ denoting the Pauli matrices in Nambu space, (ii) OP self energy, $\Sigma^{r/a}_{\delta\Delta,m-n}=[\Delta_{m-n}\tau_+ + \Delta_{n-m}^*\tau_-]$ for $m\neq n$, where $\Delta_{m-n\neq 0}=(1/T)\int (d\omega/(2\pi))e^{i(m-n)\Omega t} \delta\Delta(t)$, (iii) reservoir self-energy, $\Sigma^{r/a/<}_{\text{res.},m-n}=\zeta^2\tau_3 g_{b,m-n} \tau_3$, where $g_{b}$ is the boundary Green's function~\cite{Samanta1998}, acting only on the lead sites immediately neighbouring the reservoir ($L1/RN_R$ ), (iv) broadening self-energy $\Sigma^{r/a}_\Gamma=\pm i\Gamma/2$ and $\Sigma^{<}_\Gamma=-i\Gamma f(\omega)$, where $f(\omega)$ is the Fermi function. It aids numerical convergence, and accounts for the lifetime arising from, e.g., relaxation to the quasiparticle continuum, electron-phonon interaction, etc~\cite{Lahiri2023}. Following this prescription, the Floquet components of the current are obtained as~\cite{Lahiri2023,Cuevas1996,Gavenskythesis},
\begin{equation}
\begin{split}
I_n=&\sum_{l,m}e \int_0^\Omega \frac{d\omega}{2\pi} \mathbf{tr}\big[ \tau_3\Sigma_{\mathcal{T},LR,m+n-l} G_{R1LN_1,lm}^{<}(\omega)\\
&-(L\leftrightarrow R)\big].\end{split}\label{If}
\end{equation}

\textit{Results.--} The numerical solution to Eqs. \eqref{KDyson} and \eqref{gapeq} is presented in Fig. \ref{Fig2}. In Figs. \ref{Fig2} (a,b) we show the amplitude and phase of $\Delta_{L,m=2}$ at the first site in the left lead neighbouring the junction, such that the oscillating component of the OP is given by $2|\Delta_{L,2}(x)|\cos(\omega_Jt-\angle \Delta_{L,2}(x))$, revealing the resonant behaviour at $\omega_J=\omega_{H,L}$ characterised by a peak in the amplitude and a jump in the phase as $\omega_J$ rises across $\omega_{H,L}$. These are characteristic of a driven harmonic oscillator (c.f. Eq.~\eqref{ddelgl}), swept across its resonance~\cite{Schwarz2021}. This resonance manifests in the Floquet components of the current, shown in Fig.~\ref{Fig2} (c), where $\Delta_{0,L}$ varies while $eV$ stays fixed. The $2\omega_J$ component ($I_{m=4}$) mirrors the Higgs resonance, eventually becoming stronger than the usual $\omega_J$ component ($I_{m=2}$) if the resonance happens at a sufficiently strong gap asymmetry. We emphasise that $I_{m=4}$ does not dominate in conventional single-band SCs with \emph{time-constant} gaps~\cite{trsnote,Sellier2004,Sickinger2012,Goldobin2007,Frolov2006,Li2019,Yokoyama2014,Tanaka1997,Ng2009,Asano2003,Trimble2021,Can2021,Tummuru2022,Sun2024}. This is confirmed in the subsequent panels, showing the magnitude of $I_{m=4}$ with varying $\Delta_{0,L}$ as well as $eV$, both in the absence (Fig.~\ref{Fig2} (d)) and presence (Fig. \ref{Fig2} (e)) of the Higgs oscillations. Note that the former includes all quasiparticle-mediated processes, including the multiple Andreev reflection current, to all orders in $\mathcal{T}$. A pronounced peak in $I_{m=4}$ appears in the case of highly asymmetric junctions $\Delta_{0,L}\ll\Delta_{0,R}$ only when Higgs oscillations are present (Fig.~\ref{Fig2} (e)). In contrast, when Higgs oscillations are absent (Fig.~\ref{Fig2}(d)), the current is dominated by higher-order Josephson contributions as the junction becomes more symmetric. Finally, Fig.~\ref{Fig2}(f) shows the dependence of the $2\omega_J$ and $\omega_J$ currents on $\mathcal{T}$ for a fixed set of equilibrium gaps. We find that the former dominates only when Higgs oscillations are present, and this dominance becomes more pronounced as $\mathcal{T}$ increases.
 
\begin{figure}
\includegraphics[width=\columnwidth]{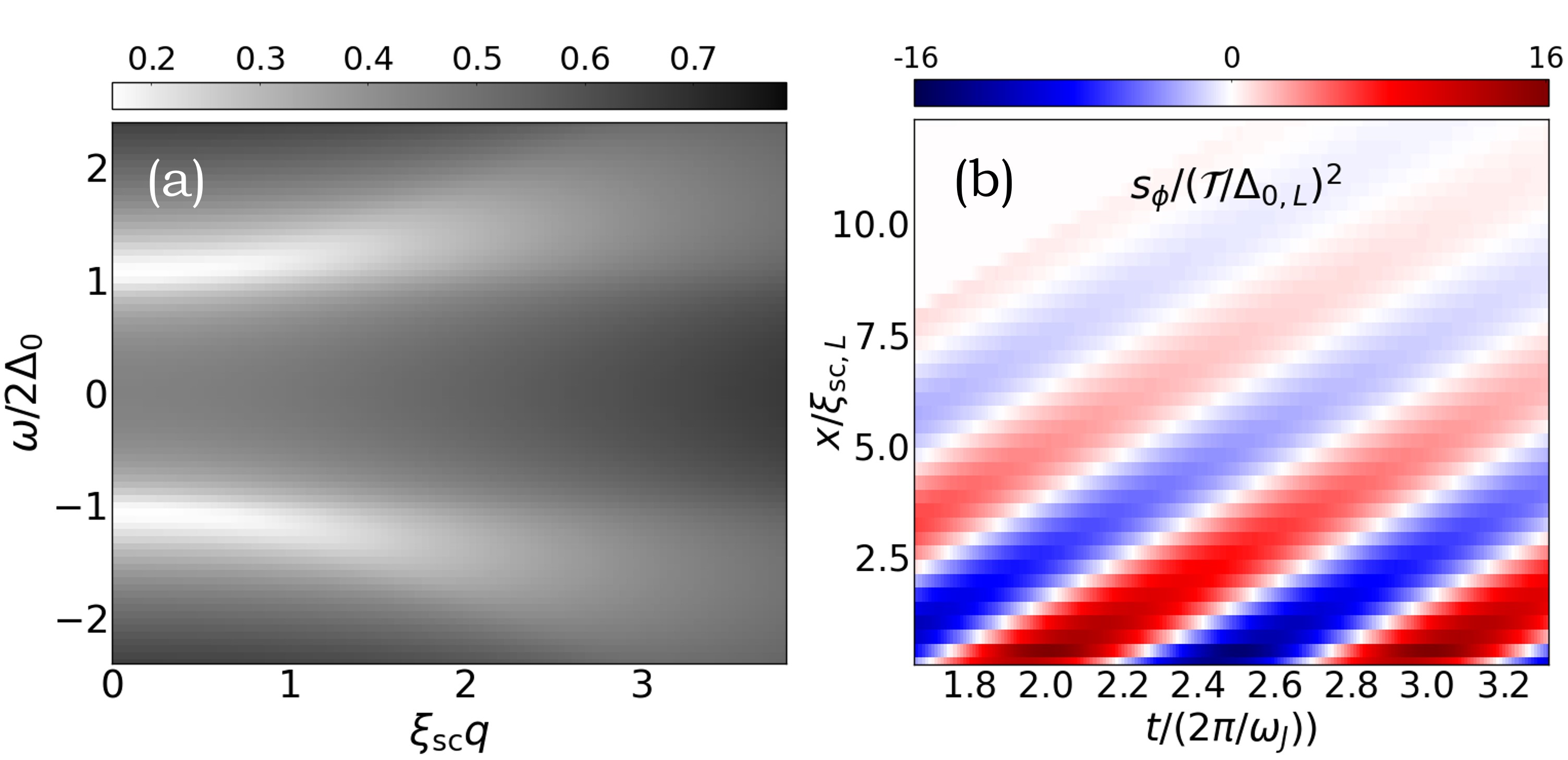}
\caption{(a) $\zeta\Re \chi_{\Delta\Delta}^{-1}(\omega,q) = \zeta(\frac{1}{g}+\frac{1}{2}\Im\chi_{\Delta\Delta,0}(\omega,q))$ for a three-dimensional s-wave SC, showing a dip at the Higgs frequency $\omega_H(q)$. We use $\zeta=10\Delta_0$, $\mu=0$, $\Gamma=0.1\Delta_0$, and $\xi_{\text{sc}}=4\zeta/(\pi\Delta_0)$ is the coherence length. (b) $s_\phi/(\mathcal{T}/\Delta_{0,L}^2)$ for $\zeta=10\Delta_{0,L}$, $\mu=0$, $\Gamma=0.25\Delta_{0,L}$ and $\Delta_{0,R}=5\Delta_{0,L}$. It oscillates at $\omega_J$, and decays into the lead $(x>0)$ over a few $\xi_{\text{sc},L}$. }
\label{Fig3} 
\end{figure} 
Finally, using a perturbative approach, we show that the Keldysh-Gorkov formulation reduces to the phenomenological model presented earlier, establishing the qualitative equivalence. Starting with the bulk equilibrium gap $\Delta_0$, we write $\Delta(t,x)=\Delta_0+\delta\Delta(t,x)$ for each lead. The corresponding self-energy is $\Sigma_{\delta\Delta}(t,x)=\delta\Delta(t,x)\tau_{x}$. The tunneling self energy is $\Sigma_{\mathcal{T},LR}(t)=\Sigma_{\mathcal{T},RL}^*(t)=(-\mathcal{T})\tau_{z}\exp(-i\phi(t)\tau_{z}/2)$. As detailed in the SM~\cite{sm}, expanding the lesser Green's functions $G_{L/R}^<$ up to $\mathcal{O}(\mathcal{T}^3)$, $G_{L/R}^<=g_{L/R}^<+\sum_{j=1}^3 (\delta G_{L/R}^<)^{(j)}$, we obtain Eq. \eqref{ddelgl} with
\begin{equation}
\begin{split}
\chi_{\Delta\Delta,L}^{-1}=&\ \frac{1}{g}-\frac{\partial\ \delta (G_{L/R}^<)^{(1)}}{\partial\ \delta\Delta_L(t,x)}\coloneqq \frac{1}{g}-i\chi_{\Delta\Delta,0,L}, \\ 
s_{\phi, L}=&\  \delta (G_{L/R}^<)^{(2)},\\
s'_{\phi,L}=&\  \frac{\partial\ \delta (G_{L/R}^<)^{(3)} }{\partial\ \delta\Delta_R(t,x)}.
\end{split} \label{couposc0}
\end{equation}
Here, $\delta (G_{L/R}^<)^{(1)}$ has one instance of $\Sigma_{\delta\Delta}$, and accounts for the intrinsic dynamics of the pairs. It defines the Higgs susceptibility. $\delta (G_{L/R}^<)^{(2)}$ has two instances of $\Sigma_{\mathcal{T}}$. It describes the effect of the radiating pairs tunneling across the junction, and defines the source term. Finally, $\delta (G_{L/R}^<)^{(3)}$ has one instance of $\Sigma_{\delta\Delta}$ and two instances of $\Sigma_{\mathcal{T}}$. It captures the cross-coupling between the OP oscillations of the two leads mediated by the tunneling pairs. We relegate the expressions for Eq. \eqref{couposc0} to the SM~\cite{sm}, focussing only on their physical behaviour here.

The OP response is governed by $\chi_{\Delta\Delta}(\omega,q)=[(1/g)+(1/2)\Im\chi_{\Delta\Delta,0}(\omega,q)-(i/2)\Re\chi_{\Delta\Delta,0}(\omega,q)]^{-1}$, where $\chi_{\Delta\Delta,0}(\omega,q)=\int\frac{d\bar{\omega}}{2\pi}\int\frac{d\bar{k}}{2\pi} \mathbf{tr}[\tau_1 (g_L(\bar{\omega}+\omega;\bar{k}+q)\tau_1 g_L(\bar{\omega};\bar{k}))^<]$ is the pair correlation. Microscopically, $\chi_{\Delta\Delta}$ describes how tunneling processes and past OP variations propagate to affect the present OP. The resulting dynamics is governed by the singularity of $\chi_{\Delta\Delta}(\omega,q)$ at the Higgs frequency $\omega_H(q)$, with $\omega_H(q=0)=2\Delta_0$. Both source and cross-coupling terms are second order in tunneling and oscillate at the Josephson frequency $2eV$, as each tunneling pair provides energy $2eV$. Hence, the OP is resonantly excited when the radiated pair energy $2eV$ matches $\omega_H$. We elaborate on the details in the SM~\cite{sm}, and show numerically obtained results for $\Re \chi_{\Delta\Delta}^{-1}(\omega,q)$ and $s_\phi$, in Fig. \ref{Fig3}.

\textit{Discussion.--} We propose the excitation and detection of the Higgs mode in SCs using the AC Josephson effect without external irradiation. In asymmetric junctions~\cite{Li2001,Ternes2006,Marchegiani2022,Ivry2014,Connolly2024,McEwen2024,Nho2025} having sufficiently different equilibrium gaps, which may be tuned experimentally~\cite{Ivry2014,Marchegiani2022,Connolly2024,McEwen2024,Nho2025}, along with high transparency, the second Josephson harmonic oscillating at twice the usual AC Josephson frequency dominates the current. As such, it may be readily observed, for instance, in the Josephson radiation~\cite{Deacon2017,Laroche2019,Liu2025}. Ideally, we seek wide specular junctions to ensure that all transport channels/subbands experience the Higgs resonance effects~\cite{Lee2023}, along with low environmental electromagnetic absorption~\cite{Holst1994,Hofheinz2011,Deacon2017,Liu2025} at frequencies comparable to the smaller superconducting gap to ensure maximal Higgs excitation. Remarkably, our results are robust to the presence of Dynes broadening, for which we have tested values as large as $\Gamma/\Delta_0\sim 0.15$, while typical experimentally relevant values are $\Gamma/\Delta_0\sim 10^{-3}-10^{-1}$ \cite{Dynes1978,Dynes1984}. Furthermore, recent studies have shown that weak paramagnetic impurities (scattering rate $1/\tau_p\ll \Delta_0$) can push the Higgs mode below the quasipaticle continuum~\cite{Li2024,Dzero2024}, thereby suppressing its decay and enhancing its signatures. 

\let\oldaddcontentsline\addcontentsline
\renewcommand{\addcontentsline}[3]{}

\begin{acknowledgements} This work was supported by the W\"urzburg-Dresden Cluster of Excellence ct.qmat, EXC2147, project-id 390858490, and the DFG (SFB 1170). We thank the Bavarian Ministry of Economic Affairs, Regional Development and Energy for financial support within the High-Tech Agenda Project ``Bausteine f\"ur das Quanten Computing auf Basis topologischer Materialien". The authors gratefully acknowledge the scientific support and HPC resources provided by the Erlangen National High Performance Computing Center (NHR@FAU) of the Friedrich-Alexander-Universität Erlangen-Nürnberg (FAU) under the NHR project b169cb. NHR funding is provided by federal and Bavarian state authorities. NHR@FAU hardware is partially funded by the German Research Foundation (DFG) – 440719683.
\end{acknowledgements}

\let\addcontentsline\oldaddcontentsline

\pagebreak
\onecolumngrid

\makeatletter

\renewcommand{\thesection}{S\arabic{section}}
\renewcommand{\theequation}{S\arabic{equation}}
\renewcommand{\thefigure}{S\arabic{figure}}
\renewcommand{\bibnumfmt}[1]{[S#1]}
\renewcommand{\citenumfont}[1]{S#1}

\newpage
\begin{large}
\begin{center}
\textbf{Supplemental Material: AC Josephson Signatures of the Superconducting Higgs Mode} 
\end{center}  
\end{large}

{\justifying
In this Supplemental Material, we present: (S1) A perturbative analysis deriving the Higgs susceptibility and source term, and establishing the qualitative equivalence between the microscopic Keldysh model which we use for our numerical calculations and the simple toy model in the introduction of the main text. (S2) Additional numerical results to show the effect of varying junction coupling, bandwidth, and device width/transverse modes.
}
\setcounter{page}{1}
\setcounter{secnumdepth}{3}
\setcounter{section}{0}
\setcounter{equation}{0}
\setcounter{figure}{0}
\setcounter{table}{0}

\tableofcontents

\newpage

\section{Perturbative analysis}\label{pertsec}
\subsection{Dyson equation for $G^<$}
\begin{equation}
\begin{split}
& (\delta G_L^<)^{(1)}=(g_L\cdot \Sigma_{\delta\Delta_L}\cdot g_L)^< =\big(g_L^r\cdot \Sigma_{\delta\Delta_L}\cdot g_L^<+g_L^<\cdot \Sigma_{\delta\Delta_L}\cdot g_L^a\big),\\
&(\delta G_L^<)^{(2)}= (g_L\cdot  \Sigma_{T,LR}\cdot g_R\cdot \Sigma_{T,RL}\cdot g_L)^< =\sum_{ \substack{l_{1..3}=\{<aa,\\ \ r<a,\ rr<\}} } g_L^{l_1}\cdot  \Sigma_{T,LR}\cdot g_R^{l_2}\cdot \Sigma_{T,RL}\cdot g_L^{l_3},\\
&(\delta G_L^<)^{(3)}=(g_L\cdot \Sigma_{T,LR}\cdot g_R\cdot \Sigma_{\delta\Delta_R}\cdot g_R \cdot \Sigma_{T,RL}\cdot g_L)^<=\sum_{ \substack{m_{1..4}=\\ \{<aaa,
\ r<aa,\\rr<a,\ rrr<\}} } g_L^{m_1}\cdot \Sigma_{T,LR}\cdot g_R^{m_2}\cdot \Sigma_{\delta\Delta_R}\cdot g_R^{m_3} \cdot \Sigma_{T,RL}\cdot g_L^{m_3},
\end{split}\label{Glessexpex}
\end{equation}
where the lowercase letter denotes the bare Green's functions, and the products are convolutions in space and time. The second equality in each line follows from the Langreth rules. 

\subsection{Coupled oscillator correspondence}
\label{couposcmicro}
From Eq. \eqref{gapeq}, $\Delta_j(t)=igF^<_{j,j}(t,t)$, we collect the first three leading terms in $\mathcal{T}$,
\begin{align}
\Delta_{0,L}(t,x)+\delta\Delta_L(t,x) =& \Re \frac{ig}{2}\mathbf{tr}[\tau_1 G_L^<(t,t;x,x)]=\Re \underbrace{\frac{ig}{2}\mathbf{tr}[\tau_1 g_L^<(t,t;x,x)]}_{=\Delta_{0,L}}+\Re \frac{ig}{2}\mathbf{tr}[\tau_1 \delta G_L^<(t,t;x,x)]\nonumber \\
\implies \frac{1}{g}\delta\Delta_L(t,x) =&\Re \frac{i}{2}\mathbf{tr}[\tau_1 (g_L\cdot \Sigma_{\delta\Delta_L}\cdot g_L)^<(t,t;x,x)]+\Re \frac{i}{2}\mathbf{tr}[\tau_1 (g_L\cdot  \Sigma_{T,LR}\cdot g_R\cdot \Sigma_{T,RL}\cdot g_L)^<(t,t;x,x)]\nonumber\\
&+\Re \frac{i}{2}\mathbf{tr}[\tau_1 (g_L\cdot \Sigma_{T,LR}\cdot g_R\cdot \Sigma_{\delta\Delta_R}\cdot g_R \cdot \Sigma_{T,RL}\cdot g_L)^<(t,t;x,x)].\quad (\text{Using Eq. }\eqref{Glessexpex})
\end{align}

Note that by considering $\Sigma_{\delta\Delta}=\delta\Delta\tau_1$ we assume a real gap as mentioned in the main text. On rearranging, we recover Eq. \eqref{ddelgl} with the microscopic version of susceptibility and sources given by Eq. \eqref{couposceq}. 
\begin{equation}
\begin{split}
\chi_{\Delta\Delta,\alpha}^{-1}(r,r')=&\frac{\delta(r-r')}{g} + \frac{\Im \mathbf{tr}[\tau_1 (g_{\alpha}{\scriptstyle(r,r')}{\cdot} \tau_1 {\cdot} g_{\alpha}{\scriptstyle(r',r)})^<]}{2},\\
s_{\phi,\alpha}(r)=&-\frac{\bigg[ \splitfrac{\Im \mathbf{tr}[\tau_1 (g_\alpha{\scriptstyle(r,r_1)}{\cdot} \Sigma_{T,\alpha\alpha'}{\scriptstyle(r_1)} }{  {\cdot} g_{\alpha'}{\scriptstyle(r_1,r_2)}{\cdot} \Sigma_{T,\alpha'\alpha}{\scriptstyle(r_2)}{\cdot} g_{\alpha}{\scriptstyle(r_2,r)})^<]} \bigg]}{2},\\
s_{\phi,\alpha}'(r,r')=&\frac{\bigg[ \splitfrac{\Im \mathbf{tr}[\tau_1 (g_{\alpha}{\scriptstyle(r,r_1)}{\cdot} \Sigma_{T,\alpha\alpha'}{\scriptstyle(r_1)} {\cdot} g_{\alpha'}{\scriptstyle(r_1,r')} }{ {\cdot} \tau_1{\cdot}  g_{\alpha'}{\scriptstyle(r',r_2)} {\cdot}\Sigma_{T,\alpha'\alpha}{\scriptstyle(r_2)}{\cdot} g_{\alpha}{\scriptstyle(r_2,r')})^<]} \bigg]}{2},
\end{split} \label{couposceq}
\end{equation}
where $\alpha,\alpha'=L/R$ denote the leads, $r\equiv (t;x)$, and $\cdot$ denotes convolution in space and time with $r_{1,2}$ integrated over. Note that $\Sigma_T\neq 0$ only at $x=0$.

In the following section, we obtain the expression for $\chi$ and elaborate on its essential features. We also obtain the behaviour of $s_\phi$, which is plotted numerically in the main text. Their features qualitatively match those obtained from the phenomenological coupled-oscillator model in the section ``Phenomenology" in the main text, thereby establishing the correspondence.

\subsection{Non-equilibrium order parameter}
Here we obtain the perturbative expression for the non-equilibrium variation in the OP. Considering a two-dimensional system (planar-junction) with Galilean invariant leads for simplicity, and assuming specular tunneling, starting with the equation for the non-equilibrium gap Eq. \eqref{ddelgl} and neglecting the cross-terms, we have,
\begin{align}
&\frac{\delta\Delta_L\big(t;\{x,y\}\big)}{g}=-\frac{1}{2}\Im \int_{t_1}\sum_{x_1,y_1}\mathbf{tr}\big[\tau_1 g_L(t,t_1;\{x,y\},\{x_1,y_1\})\tau_1 g_L(t_1,t;\{x_1,y_1\},\{x,y\})\big]^<\delta\Delta\big(t_1;\{x_1,y_1\}\big)\nonumber\\
&-\frac{1}{2}\Im \iint_{t_1,t_2}\sum_{y_1,y_2}\mathbf{tr}\big[ \tau_1 g_L\big(t,t_1;\{x,y\},\{0,y_1\}\big)\Sigma_{T,LR}(t_1) g_R\big(t_1,t_2;\{0,y_1\},\{0,y_2\}\big)\Sigma_{T;RL}(t_2)g_L\big(t_2,t;\{0,y_2\},\{x,y\}\big) \big]^<.
\end{align}
On Fourier transforming,
\begin{align}
&\frac{\sum_{\bf{q}}}{N_x N_y}\int \frac{d\omega}{(2\pi)}\frac{\delta\Delta_L(\omega;{\bf{q}})}{g}e^{-i\omega t+i{\bf{q}}\cdot {\bf{r}}}=-\frac{1}{2}\Im \frac{\sum_{\bf{q},\bf{k}_2}}{(N_x N_y)^2}\int \frac{d\omega}{(2\pi)}\frac{d\omega_2}{(2\pi)} \mathbf{tr}\big[\tau_1 g_L(\omega_2+\omega;{\bf{k_2}+\bf{q}})\tau_1 g_L(\omega_2;{\bf{k_2}})\big]^<\delta\Delta(\omega;{\bf{q}})e^{-i\omega t+i{\bf{q}}\cdot {\bf{r}}}\nonumber\\
&-\frac{1}{2}\Im \frac{\sum_{\substack{ k_{1,x},k_{2,x}, \\ k_{3,x},k_{3,y} }}}{N_x^3 N_y}\int\frac{d\Omega_1 d\Omega_2d\omega_3}{(2\pi)^3}\mathbf{tr}\big[ \tau_1 g_L(\omega_3+\Omega_1+\Omega_2;k_{1,x},k_{3,y})\Sigma_{T,LR}(\Omega_1) g_R(\omega_3+\Omega_2;k_{2,x},k_{3,y})\Sigma_{T;RL}(\Omega_2)g_L(\omega_3;k_{3,x},k_{3,y}) \big]^<\nonumber\\ &\hspace*{4.96cm} e^{-i(\Omega_1+\Omega_2) t+i(k_{1,x}-k_{3,x})x+ik_{3,y}y},
\end{align}
where we have considered specular tunneling, and $\Sigma_{T,LR/RL}(\Omega)=-\mathcal{T}\begin{bmatrix}\delta(\Omega\mp eV) & 0\\ 0 & -\delta(\Omega\pm eV) \end{bmatrix}$.

Assuming transverse homogeneity, i.e., only the $q_y=0$ component of the OP is non-zero, we obtain,
\begin{align}
\delta\Delta(\omega;q_x,0)=&\chi_{\Delta\Delta,L}(\omega;q_x,0)\frac{s_{\phi,L}(\omega;q_x,0)-s^*_{\phi,L}(-\omega;-q_x,0)}{2i},\label{deldelta}
\end{align}
where 
\begin{align}
\chi_{\Delta\Delta,L}(\omega;q_x,0)=&\frac{1}{\frac{1}{g}+\frac{1}{2}\frac{\chi_{\Delta\Delta,0,L}(\omega;q_x,0)-\chi_{\Delta\Delta,0,L}^*(-\omega;-q_x,0)}{2i}},\nonumber\\
 \chi_{\Delta\Delta,0,L}(\omega;q_x,0)=&\int\frac{d\bar{\omega}}{2\pi}\int\frac{d\mathbf{k}}{(2\pi)^2} \mathbf{tr}[\tau_1 (g_L(\bar{\omega}+\omega;\mathbf{k}+q_x)\tau_1 g_L(\bar{\omega};\mathbf{k})))^<] ,
\end{align}
and
\begin{align}
s_{\phi,L}(\omega;q_x,0)=&  -\frac{1}{2}\frac{\sum_{k_{2,x},k_{3,x}}}{N_x^2}\int\frac{d\Omega_2d\omega_3}{(2\pi)^2}\mathbf{tr}\big[ \tau_1 g_L(\omega_3+\omega;k_{3,x}+q_x,0)\Sigma_{T,LR}(\omega-\Omega_2) g_R(\omega_3+\Omega_2;k_{2,x},0)\Sigma_{T;RL}(\Omega_2)\nonumber\\
&\hspace*{4.7cm}g_L(\omega_3;k_{3,x},0) \big]^<.\label{swq}
\end{align}
Later, we convert the momentum sums $\frac{\sum_{k_x}}{N_x}\to \int\frac{dk_x}{2\pi}$, where $k_x$ is the dimensionless product of the wavevector and the lattice constant. The source term in the numerator generates oscillations at $\omega_J$ (along with a DC renormalisation near the junction), with the magnitude of the OP governed by the singularities (corresponding to the Higgs mode in our case) of the susceptibility in the denominator. We show in the section on susceptibility that at the Higgs resonance, $\chi_{\Delta\Delta}\sim\sqrt{\Delta_{0,L}/\Gamma}$, thus showing that $\delta\Delta_L\sim\sqrt{\Delta_{0,L}/\Gamma}$ at the Higgs resonance. Thus, the Higgs-enhanced $2\omega_J$ AC Josephson effect can be made arbitrarily strong with decreasing $\Gamma$.

\subsection{Susceptibility}
Here we look at the susceptibility which governs the magnitude of the OP response for any given frequency and wavevector. 
The susceptibility is written as,
\begin{align}
\chi_{\Delta\Delta}(\omega,q_x)=&\frac{1}{\frac{1}{g}+\frac{1}{2}\frac{\chi_{\Delta\Delta,0}(\omega,q_x)-\chi_{\Delta\Delta,0}^*(-\omega,-q_x)}{2i}}=\frac{1}{\frac{1}{g}+\frac{1}{2}\Im\chi_{\Delta\Delta,0}(\omega,q_x)-\frac{i}{2}\Re\chi_{\Delta\Delta,0}(\omega,q_x)},\label{chif}\\
\chi_{\Delta\Delta,0}(\omega,q_x)=&\int\frac{d\bar{\omega}}{2\pi}\int\frac{d\mathbf{k}}{(2\pi)^2} \mathbf{tr}[\tau_1 (g_L(\bar{\omega}+\omega;\mathbf{k}+q_x)\tau_1 g_L(\bar{\omega};\mathbf{k})))^<].\label{chi0}
\end{align}
In the second equality of Eq. \eqref{chif}, we need to substitute the BCS coupling constant $g$ by obtaining its value from the zero-temperature equilibrium gap equation,
\begin{align}
\Delta_0=&\Re \frac{ig}{2}\mathbf{tr}[\tau_1 g_L^<(t,t;x,x)]=\Re ig\int\frac{d\omega}{2\pi}\frac{d\mathbf{k}}{(2\pi)^2} i \underbrace{\frac{\mathbf{tr}[\tau_1 (-2\Im g^r)]}{2}}_{A_s(\omega,k)}f(\omega).
\end{align}
where $k=|\mathbf{k}|$. Using the retarded Green's function,
\begin{align}
g^r(\omega,k) =& \int \frac{d\omega'}{2\pi} \frac{A(\omega',k)}{\omega-\omega'+i\Gamma},\quad g^<(\omega,k)=-(g^r(\omega,k)-g^a(\omega,k))f(\omega),\\
A(\omega,k) =& 2\pi \begin{bmatrix}
u_k^2\delta(\omega-E_k)+v_k^2\delta(\omega+E_k) & u_kv_k\delta(\omega-E_k)-u_kv_k\delta(\omega+E_k)\\
u_kv_k\delta(\omega-E_k)-u_kv_k\delta(\omega+E_k) & v_k^2\delta(\omega-E_k)+u_k^2\delta(\omega+E_k)
\end{bmatrix},\quad \left(\begin{matrix} u_k(v_k)=\sqrt{\frac{1}{2}+(-)\frac{\xi_k}{2E_k}} \\ E_k=\sqrt{\xi_k^2+\Delta_0^2}\end{matrix}\right),
\end{align}
where $\xi_k=\epsilon_k-\mu=-2\zeta\cos(k)-\mu$. Thus, we obtain,
\begin{align}
\frac{1}{g}=&\frac{\nu}{2} \int_{-\infty}^0 d\xi \frac{1}{\sqrt{\Delta_0^2+\xi^2}}\frac{2\tan^{-1}\big(\frac{\sqrt{\Delta_0^2+\xi^2}}{\Gamma}\big)}{2\pi},\label{1bggap}
\end{align}
where we have used the wideband limit with a constant density of states $\nu$ at the Fermi level. 

\begin{figure}[!htb]
\includegraphics[width=2.5in]{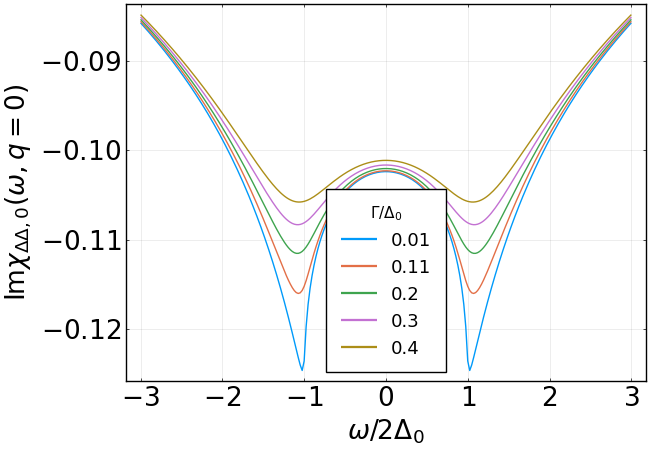}\llap{\parbox[b]{4.5in}{(a)\\\rule{0ex}{1.8in}}}\hspace*{10mm}
\includegraphics[width=2.5in]{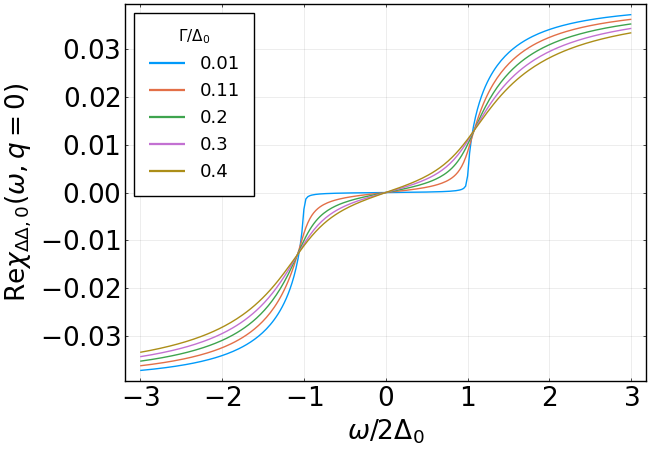}\llap{\parbox[b]{4.5in}{(b)\\\rule{0ex}{1.8in}}}
\\
\includegraphics[width=2.5in]{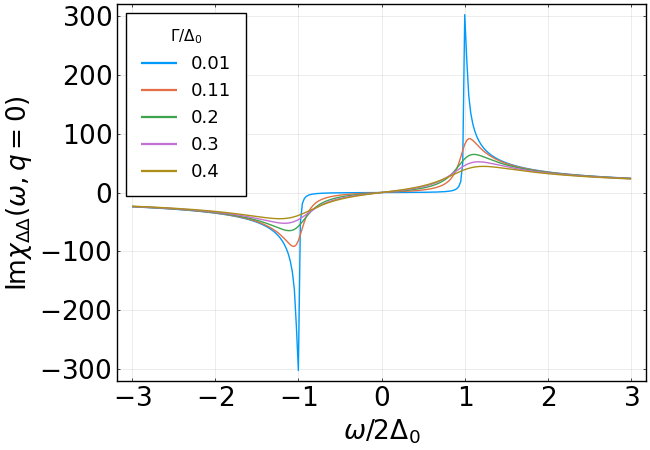}\llap{\parbox[b]{4.5in}{(c)\\\rule{0ex}{1.8in}}}\hspace*{10mm}
\includegraphics[width=2.5in]{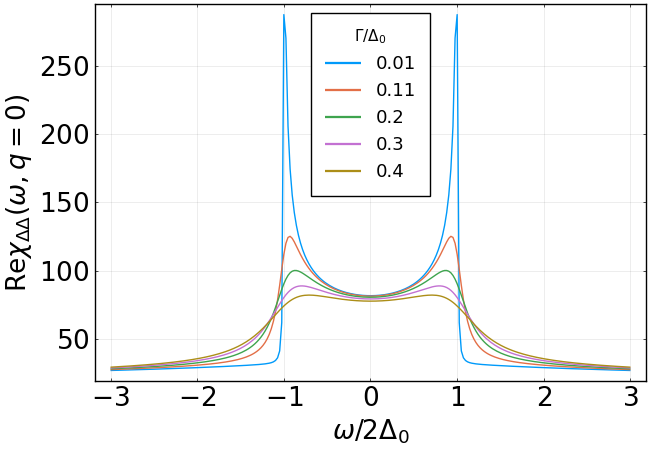}\llap{\parbox[b]{4.5in}{(d)\\\rule{0ex}{1.8in}}}\\
\includegraphics[width=2.5in]{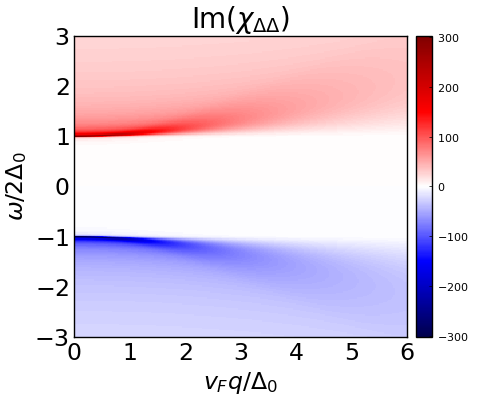}\llap{\parbox[b]{4.5in}{(e)\\\rule{0ex}{1.9in}}}\hspace*{10mm}
\includegraphics[width=2.5in]{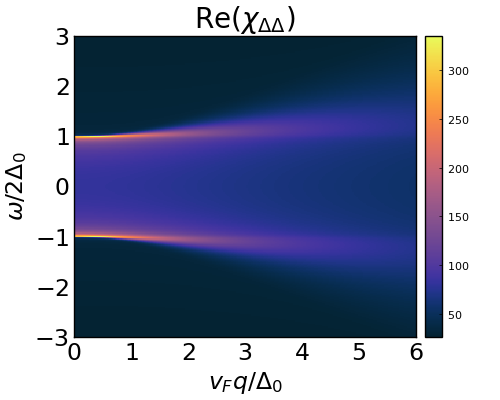}\llap{\parbox[b]{4.5in}{(f)\\\rule{0ex}{1.9in}}}
\caption{Pair susceptibility in a system with $\zeta=20\Delta_0$, $\mu=0$ (half-filled). (a,b) show $\chi_{\Delta\Delta,0}(\omega,q)$ as a function of frequency $\omega$ for $q=0$, for varying $\Gamma$. The $q=0$ Higgs mode occurs at $\omega/(2\Delta_0)=\pm 1$. (c,d) show the full susceptibility $\chi_{\Delta\Delta}(\omega,q=0)$, for varying $\Gamma$. The singularity in $\chi_{\Delta\Delta}$ at $\omega=\pm 2\Delta_0$ gets stronger with decreasing $\Gamma$. (e-f) show the full susceptibility $\chi_{\Delta\Delta}(\omega,q=\xi/v_F)$ for $\Gamma/\Delta_0=0.1$, revealing the Higgs dispersion. We state $q$ in terms of the single-particle dispersion $\xi=v_F q$, linearised near the Fermi energy with the Fermi velocity $v_F$.}
\label{FigS4} 
\end{figure}

In the expression for $\chi_{\Delta\Delta,0}$, the lesser component of the product of Green's functions is obtained using the Langreth rules. $\chi_{\Delta\Delta,0}$ is shown in Fig. \ref{FigS4}(a-b), revealing the Higgs mode dispersion. Its $q=0$ expression can be analytically derived. 
We define $\theta$ as the angle between the internally-summed wavevector $\mathbf{k}$ and the external $q$, $E_k=\sqrt{\xi_k^2+\Delta_0^2}\coloneqq \sqrt{\xi^2+\Delta_0^2}$ where $\xi=\xi_k=v_F(k-k_F)$ with $k=|\mathbf{k}|$, and $E_{k+q_x}=\sqrt{\xi_{k+q_x}^2+\Delta_0^2}\approx \sqrt{(\xi+v_Fq_xx)^2+\Delta_0^2}$ where $x=\cos(\theta)$. Here we expect the dominant contribution to come from $q_x\ll k_F$ and $k\sim k_F$, with $k_F\approx 2\zeta/v_F$ being the Fermi wavevector. Hence we approximate the otherwise energy-dependent density of states with its value at the Fermi level. Thus, we get
\begin{align}
 \chi_{\Delta\Delta,0}(\omega,q_x)=\nu \int_0^\infty d\xi \int_{-1}^1 dx \int_{-\infty}^\infty i \bigg[ &\frac{u_k'^2 u_k^2}{(\omega-E_k-E_k'+i 2\Gamma)} + \frac{v_k'^2 v_k^2}{(\omega-E_k-E_k'+i 2\Gamma)} - \frac{2 u_k' v_k' u_k v_k}{(\omega-E_k-E_k'+i 2\Gamma)}\nonumber\\
 & + \frac{v_k'^2 u_k^2}{(\omega-E_k+E_k'+i 2\Gamma)} +\frac{ u_k'^2 v_k^2}{(\omega-E_k+E_k'+i 2\Gamma)} + \frac{2 u_k' v_k' u_k v_k}{(\omega-E_k+E_k'+i 2\Gamma)} \nonumber\\ 
 & + \frac{v_k'^2 v_k^2}{(-\omega-E_k-E_k'-i 2\Gamma)} +\frac{ u_k'^2 u_k^2}{(-\omega-E_k-E_k'-i 2\Gamma)} - \frac{2 u_k' v_k' u_k v_k}{(-\omega-E_k-E_k'-i 2\Gamma)} \nonumber\\
 & + \frac{v_k'^2 u_k^2}{(-\omega+E_k-E_k'-i 2\Gamma)} + \frac{u_k'^2 v_k^2}{(-\omega+E_k-E_k'-i 2\Gamma)} + \frac{2 u_k' v_k' u_k v_k}{(-\omega+E_k-E_k'-i 2\Gamma)} \bigg]
\end{align}
where $E_k'=E_{|\mathbf{k}+q_x|}$ and $u_k'(v_k')=u_{|\mathbf{k}+q_x|}(v_{|\mathbf{k}+q_x|})$. Note that $\chi_{\Delta\Delta}$ depends only on the magnitude of $q_x$ in isotropic s-wave SCs, which is what we consider in this work. For $q_x=0$, there exists a compact analytical expression,
\begin{align}
\chi_{\Delta\Delta,0}(\omega,q_x=0)=&i\nu \int_0^\infty d\xi \int_{-1}^1 dx \frac{4\xi^2}{\sqrt{\xi^2+\Delta_0^2}((\omega+i2\Gamma)^2-4\xi^2-4\Delta_0^2)}\nonumber\\
=& -i\frac{2}{g} + i2\nu\frac{\tanh^{-1}\left(\sqrt{1+\frac{1}{\left(\frac{\omega}{2\Delta_0}\right)^2-1+i\left(\frac{\omega}{2\Delta_0}\right)\left(\frac{\Gamma}{\Delta_0}\right)}}\right)}{\sqrt{1+\frac{1}{\left(\frac{\omega}{2\Delta_0}\right)^2-1+i\left(\frac{\omega}{2\Delta_0}\right)\left(\frac{\Gamma}{\Delta_0}\right)}}}.\label{chi0Higgsq0f}
\end{align}
The first term is $i(2/g)$, which is evident from Eq. \eqref{1bggap}. It eventually cancels the logarithmic UV divergence in the denominator of $\chi_{\Delta\Delta}$.  $\Re \chi_{\Delta\Delta,0}(\omega,q=0)$, which governs the lifetime of the Higgs mode (see Eq. \eqref{chif}), is bounded. It is exponentially small for $|\Omega|<2\Delta_0$ and satisfies $\Re \chi_{\Delta\Delta,0}(\omega\gg 2\Delta_0,q=0)=\nu(\pi/2)$ and $\Re \chi_{\Delta\Delta,0}(\omega\ll -2\Delta_0,q=0)=-\nu(\pi/2)$. $(1/2)\Im \chi_{\Delta\Delta,0}(\omega,q=0)+(1/g)$ shows a dip at $\omega=\pm 2\Delta_0$, touching zero for $\Gamma/\Delta_0\to 0$.
\begin{align}
\chi_{\Delta\Delta,0}(\omega=2\Delta_0,q_x=0) \approx & i\bigg(-\frac{2}{g} + \nu\frac{\pi}{2}\sqrt{\frac{\Gamma}{\Delta_0}}\bigg) + \bigg(\nu\frac{\pi}{2}\sqrt{\frac{2\Gamma}{\Delta_0}} - 4\nu\frac{\Gamma}{\Delta_0}\bigg).\label{chi0Higgsq0}
\end{align}
Hence, from Eqs. \eqref{chif}, \eqref{chi0Higgsq0}, and \eqref{1bggap}, $\chi_{\Delta\Delta}$ derives its singular enhancement from that of $\chi_{\Delta\Delta,0}$ at the Higgs resonance,
\begin{align}
\chi_{\Delta\Delta}(\omega=2\Delta_0,q_x=0)\approx &\frac{1}{ \nu\frac{\pi}{4}\sqrt{\frac{\Gamma}{\Delta_0}} - i\big(\nu\frac{\pi}{4}\sqrt{\frac{\Gamma}{\Delta_0}} - 2\nu\frac{\Gamma}{\Delta_0}\big)}
\end{align}

\subsection{Source}
Considering a two-dimensional system (planar junction), from Eq. \eqref{couposceq}, the expression for the source $s_\phi$ in time and space domain is given by,
\begin{align}
&s_{\phi}\big(t;\{x,y\}\big)\nonumber\\
&=-\frac{1}{2}\Im \iint_{t_1,t_2}\sum_{y_1,y_2}\mathbf{tr}\big[ \tau_1 g_L\big(t,t_1;\{x,y\},\{0,y_1\}\big)\Sigma_{T,LR}(t_1) g_R\big(t_1,t_2;\{0,y_1\},\{0,y_2\}\big)\Sigma_{T;RL}(t_2)g_L\big(t_2,t;\{0,y_2\},\{x,y\}\big) \big]^<.
\end{align}
On Fourier transforming, we obtain $(s_{\phi,L}(\omega;q_x,0)-s^*_{\phi,L}(-\omega;-q_x,0))/(2i)$ with $s_{\phi,L}(\omega;q_x,0)$ given by Eq.~\eqref{swq}. The tunneling self energy $\Sigma_{T,LR/RL}(\Omega)=-\mathcal{T}\begin{bmatrix}\delta(\Omega\mp eV) & 0\\ 0 & -\delta(\Omega\pm eV) \end{bmatrix}$. Due to this form for $\Sigma_{T,LR/RL}(\Omega)$, $s_\phi$ only has terms which oscillate with frequency $\pm\omega_J=\pm 2eV$
\begin{align}
s_{\phi}(\omega;q_x,0)=&\delta(\omega\mp 2eV)\frac{1}{2} \frac{\sum_{\substack{ k_{x}, p_{x}}}}{N_x^2}\int\frac{d\Omega}{(2\pi)} \big(g_{L,11/22}(\Omega\pm 2eV;k_{x}+q_x,0)g_{R,12}(\Omega\pm eV;p_{x},0) g_{L,22/11}(\Omega;k_{x},0)\nonumber\\ 
&\hspace*{4.08cm}+g_{L,12}(\Omega\pm 2eV;k_{x}+q_x,0)g_{R,12}(\Omega\pm eV;p_{x},0) g_{L,12}(\Omega;k_{x},0)\big)^<,
\end{align}
and a DC term. Note that this DC component is neglected in the phenomenological model in the main text.

\subsection{Current}
%

Here we derive the leading order change in the current due to the Higgs oscillations of the OP.
\begin{align}
I(t)=&-2e\Re\ \mathbf{tr}[\tau_3\Sigma^T_{LR} \underbrace{(g_R\Sigma^T_{RL}g_L)^{<}}_{(\delta G_{R,L}^<)^{(1)}}(t,t)]=-2e h_0(t),
\end{align}
where
\begin{align}
h_0(t)=& \Re \int dt_1\mathbf{tr}\Big[\tau_3\Sigma^T_{LR}(t) \big(g_R(t-t_1,0-0)\Sigma^T_{RL}(t_1)g_L(t_1-t,0-0)\big)^{<}\Big]. \label{IT2}
\end{align}
In order to align with the simple approximation employed in the introduction of the main text, we have to consider the ``adiabatic" approximation wherein the time-scale associated with the voltage/Josephson phase, $\sim1/eV$, is much larger than the time-scale for the oscillations and decay of the Green's functions~\cite{sSchulp1978}. For a subgap ($eV<2\Delta$) DC voltage bias with $\phi(t)=2eVt$, this yields,
\begin{align}
I=&J_1\Delta_{0,R}\Delta_{0,L}\sin(2eVt)+J_2\Delta_{0,R}\Delta_{0,L}\cos(2eVt).
\end{align}
Note the cosine term, which is typically omitted in the simple version as it is typically not relevant for $eV<2\Delta$, and it doesn't change our conclusions. The coefficient $J_1\Delta_{0,R}\Delta_{0,L}$ yields the critical current.

The change in the current due to the Higgs oscillations is subsequently obtained as,
\begin{align}
\delta I(t)=&-2e\Re\ \mathbf{tr}[\tau_3\Sigma^T_{LR} \underbrace{(g_R\Sigma^T_{RL}g_L\Sigma_{\delta\Delta_L}g_L)^{<}}_{(\delta G_{R,L}^<)^{(2)}}(t,t)]=-2e \int_{-\infty}^t dt' h(t,t';x') \delta\Delta_L(t',x'), \label{IT4}
\end{align}
where,
\begin{align}
h(t,t';x')=&\frac{\partial h_0(t)}{\partial \Delta_L(t',x')}\bigg\vert_{\Delta_L(t',x')=\Delta_{0,L}}\\
=&\Re \int dt_1\mathbf{tr}\Big[\tau_3\Sigma^T_{LR}(t) \big(g_R(t-t_1,0-0)\Sigma^T_{RL}(t_1)g_L(t_1-t',0-x')\tau_ 1 g_L(t'-t,x'-0)\big)^{<}\Big].
\end{align}
The function $h$ inherits the spatio-temporal decay from the Green's functions. That is, as a function of $t-t'$, it decays over a time-scale governed by $1/\Gamma$, whereas as a function of $x'$, it decays over a length-scale governed by $\xi_{\text{sc},L}$, the superconducting coherence of the left lead. Physically speaking, the current at the junction depends on the OP variation $\delta\Delta(t';x')$ in the region of space-time limited by these.

Within the adiabatic approximation, using the derivative form specified above and assuming that $\delta\Delta$ is peaked near the junction, we find
\begin{align}
\delta I\sim &J_1\Delta_{0,R}\delta\Delta_{L}\sin(2eVt)+J_2\Delta_{0,R}\delta\Delta_{L}\cos(2eVt),
\end{align}
which is the expression employed in the introduction. The cosine term is neglected in the main text for reasons mentioned above.

\section{Numerical analysis}
\subsection{Equilibrium gap}
\begin{figure*}[!htb]
\includegraphics[width=2.7in]{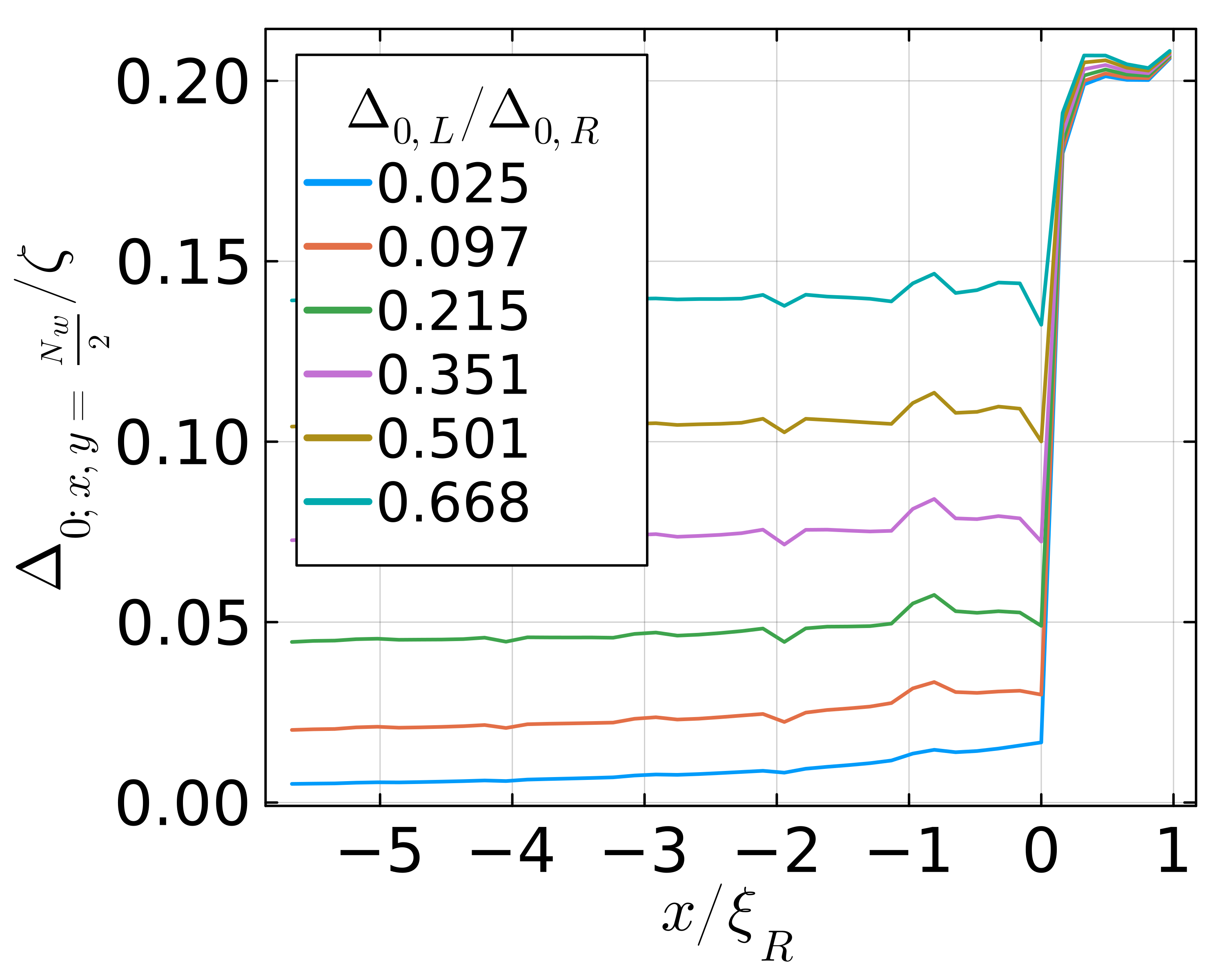}\caption{Numerically obtained equilibrium gap at the $y=6^{\text{th}}$ site. We use the same parameters as in Fig.~\ref{Fig2}. The gap in the left lead is slightly larger near the junction due to the inverse proximity effect, and stabilises to its bulk value over a distance of the order of the coherence length $\xi_L\sim 1/\Delta_{0,L}$ ($\xi_L\neq\xi_R$).}
\label{FigS0m} 
\end{figure*}
We present in Fig.~\ref{FigS0m} the equilibrium gap for varying BCS coupling in the left lead, while keeping it fixed in the right lead. The gap in the left lead is slightly enhanced near the junction by the inverse proximity effect.

\subsection{Transverse profile of the OP}
\begin{figure*}[!htb]
\includegraphics[width=2.3in]{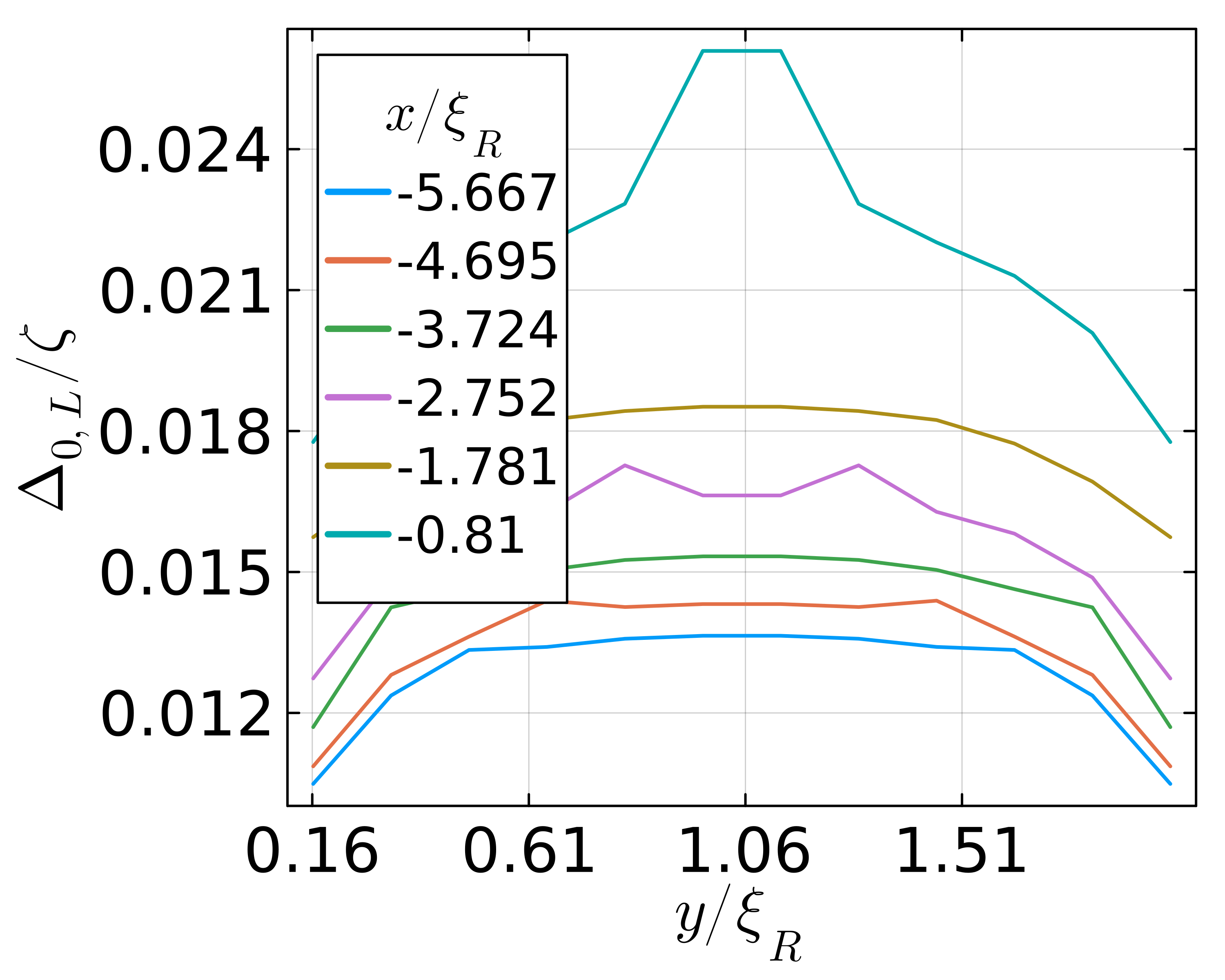}\llap{\parbox[b]{0.8in}{(a)\\\rule{0ex}{1.6in}}}\hspace*{5mm}
\includegraphics[width=2.3in]{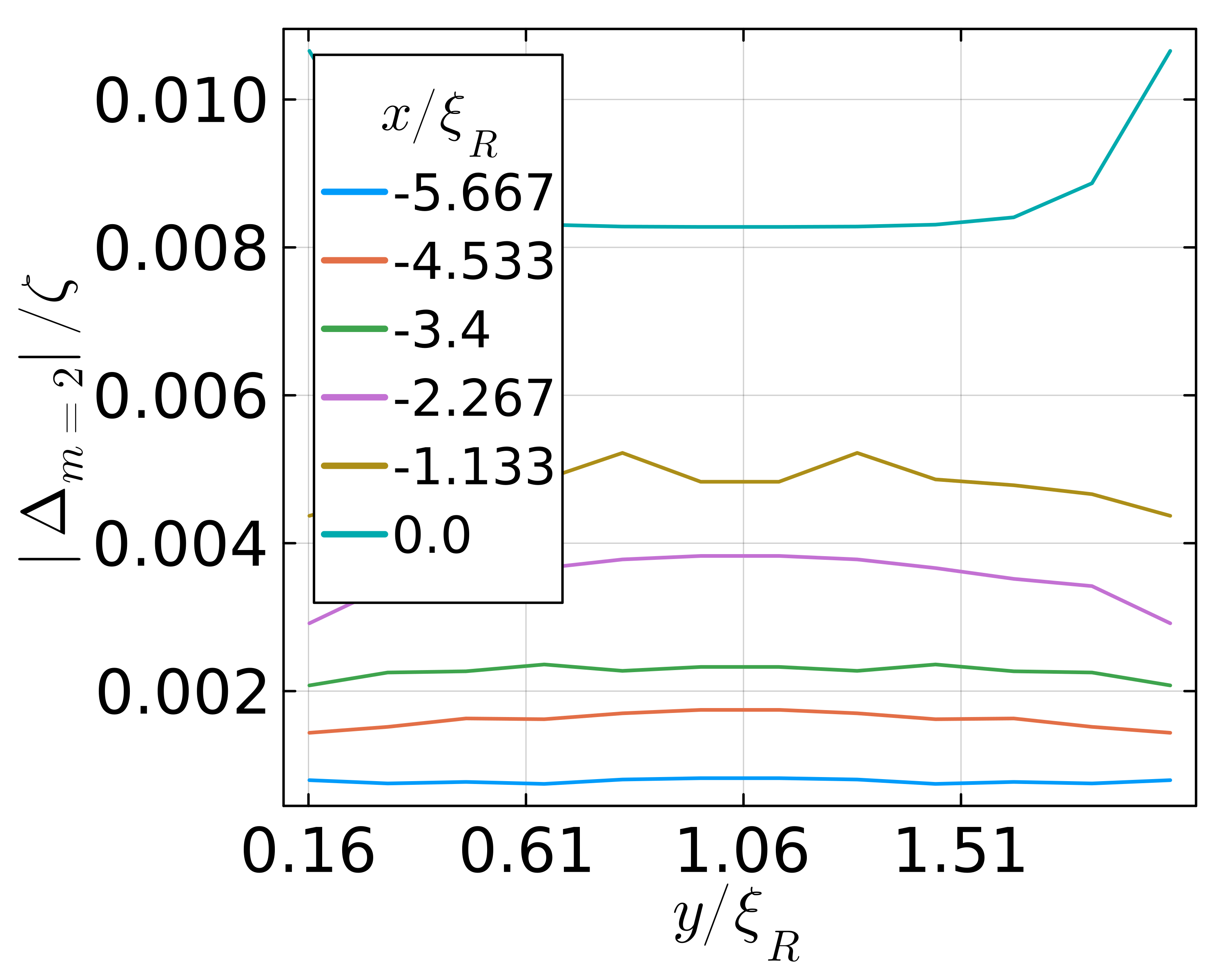}\llap{\parbox[b]{0.8in}{(b)\\\rule{0ex}{1.6in}}}
\caption{Numerically obtained equilibrium gap and non-equilibrium OP, showing the variations along the transverse direction. (a) $\Delta_{0,L}(x,y)/\zeta$ for $\Delta_{0,L}/\zeta\equiv \Delta_{0,L}(x=-\infty,y)/\zeta=0.014$, (b) $|\delta\Delta_{L,m=2}|/\zeta$ for the same equilibirum gap and $eV/\zeta=0.02$. The remaining parameters are the same as in Fig.~\ref{Fig2}.}
\label{FigS0} 
\end{figure*}
As shown in Fig.~\ref{FigS0}, we find that the OP seemingly exhibits transverse homogeneity within the bulk, with small variations mostly near the edges.

\subsection{Three dimensional junctions: Varying number of transverse modes/sub-bands}
In this section we model three dimensional Josephon junctions assuming a uniform OP along the transverse directions. In this case, the system splits into disjoint transverse subbands with different chemical potentials, determined by the subband energies. 

\begin{figure*}[!htb]
\includegraphics[width=2.25in]{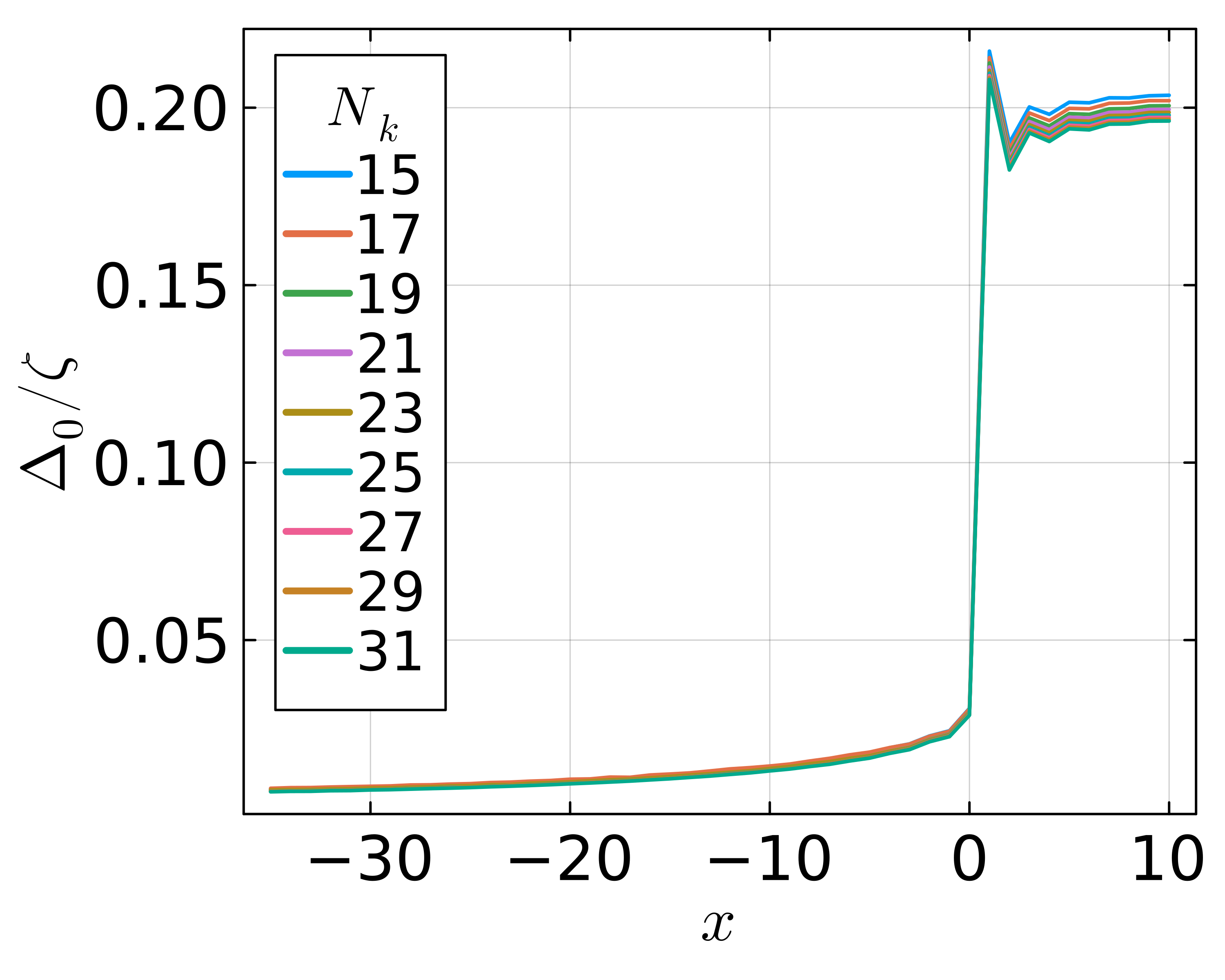}\llap{\parbox[b]{1.85in}{(a)\\\rule{0ex}{1.58in}}}\hspace*{5mm}
\includegraphics[width=2.25in]{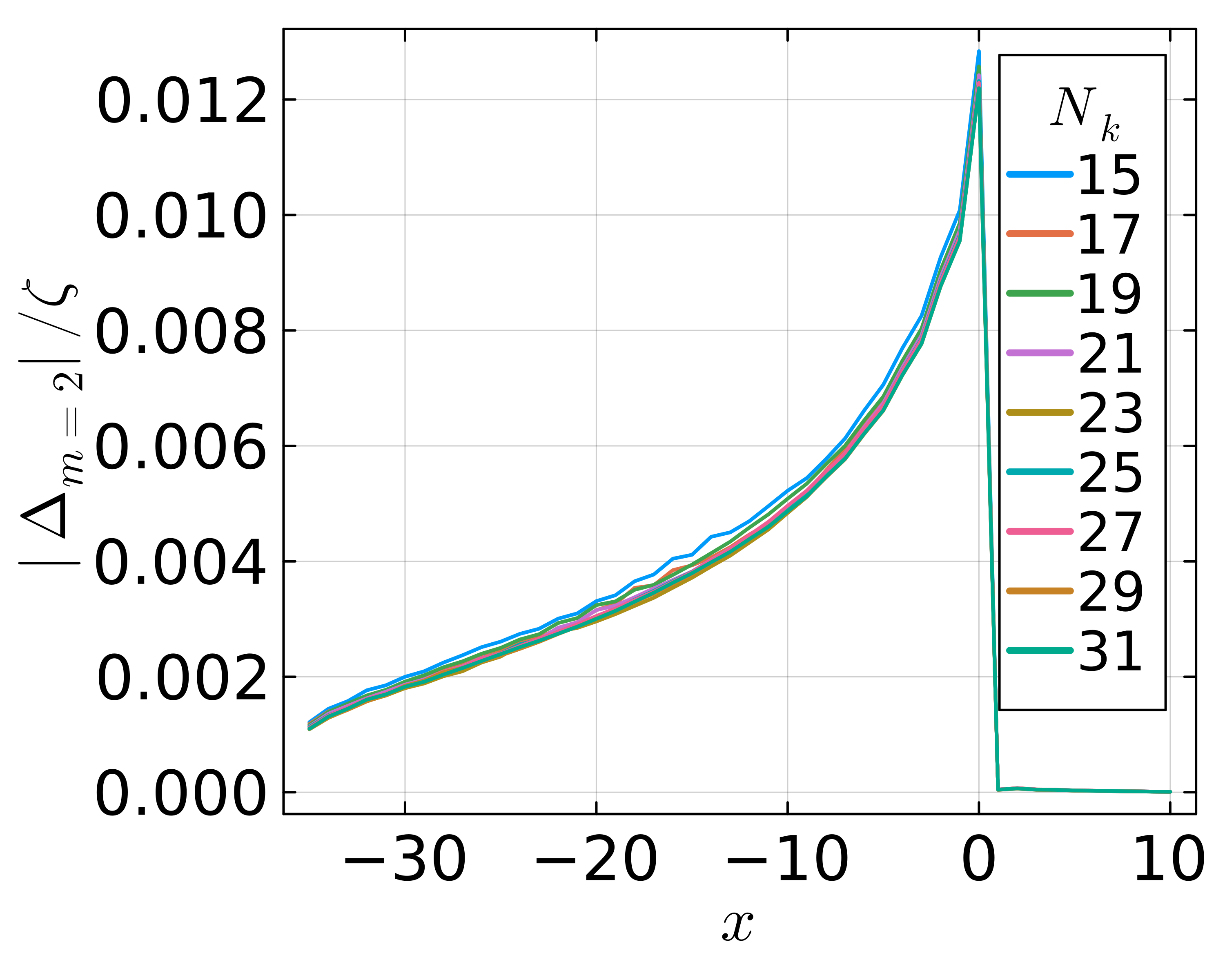}\llap{\parbox[b]{1.85in}{(b)\\\rule{0ex}{1.58in}}}
\includegraphics[width=2.25in]{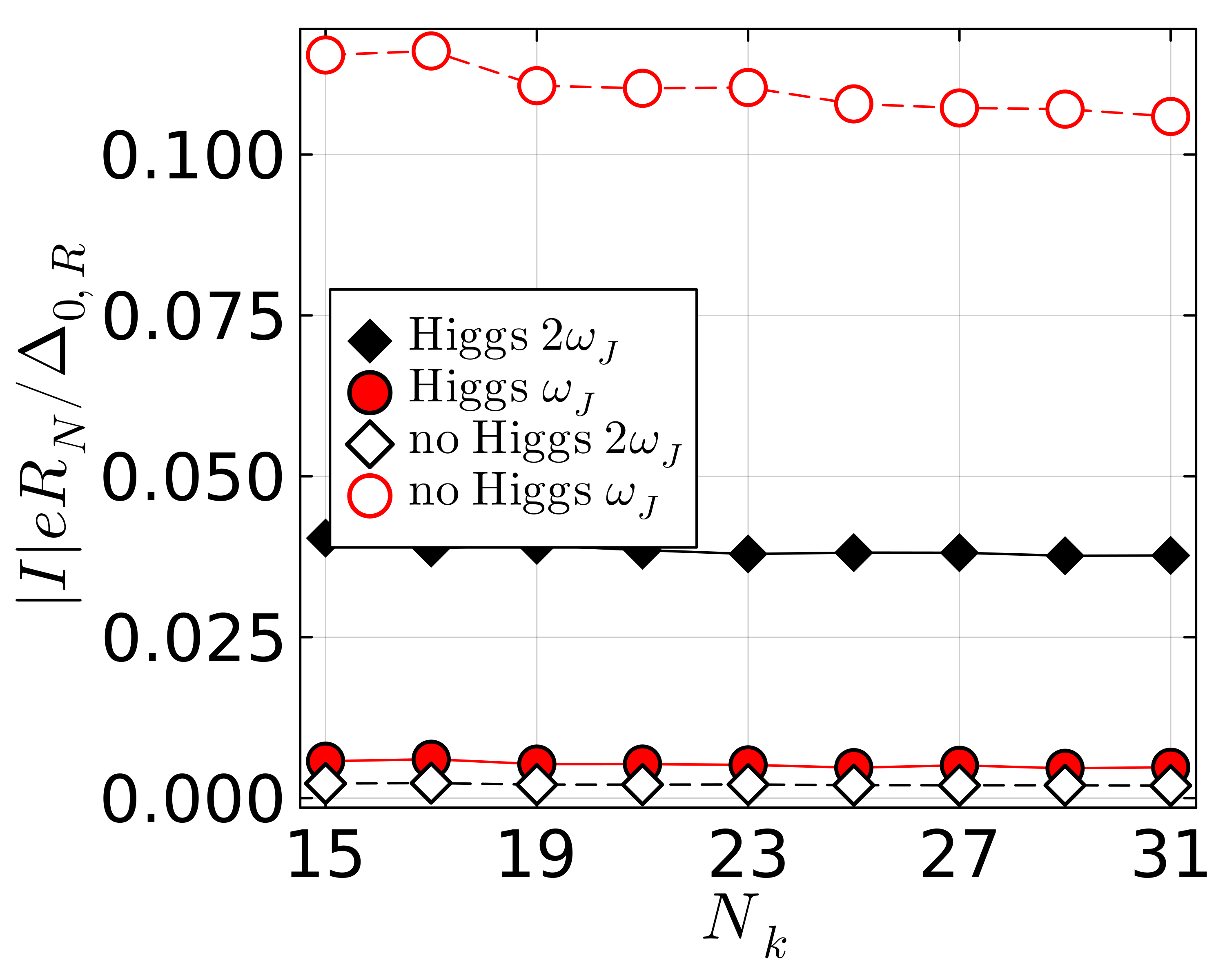}\llap{\parbox[b]{3.7in}{(c)\\\rule{0ex}{1.58in}}}
\caption{Numerically obtained results for varying number of transverse modes $N_k$, for the same BCS coupling constant, and hence, same equilibrium gaps. We consider $N_L=36$, $N_R=10$, $\zeta=5$ (bandwidth $12\zeta=60$), and $\mathcal{T}/\zeta=0.4$. (a) The fixed equilibrium gaps, with $\Delta_{0,L}/\Delta_{0,R}\approx 0.04$. (b) $\Delta_{m=2}$ at the site on the left lead immediately neighbouring the junction. It remains visibly unchanged with increasing $N_k$. (c) $I_4$ (black diamonds) and $I_2$ (red circles) are shown as functions of $N_k$, both with (solid markers) and without (empty markers and dashed lines) Higgs oscillations, for $eV/\zeta = 0.1$ ($\Delta_{0,L}/\zeta\approx 0.04$). $I_4$ exceeds $I_2$, and the trends get saturated for large $N_k$. Note that $R_N$ is numerically calculated, and it decreases with increases $N_k$. }
\label{FigS3} 
\end{figure*}

In Fig. \ref{FigS3}, we present results with varying number of transverse modes $N_k$, corresponding to varying widths of the planar junction. We find that with increasing device width, the difference in strengths of the $2\omega_J$ and $\omega_J$ Josephson currents are accentuated, resulting from the contribution from an extensively scaling number of subbands. Crucially, the order of dominance is not affected, i.e., if the $2\omega_J$ current is stronger in the few-channel/subband limit, then it remains stronger even on increasing the device width.

Considering a three-dimensional junction, considering $N_y=N_z=N_k$ bands/channels along the transverse $y-$ and $z-$directions (perpendicular to the longitudinal $x-$direction), the Hamiltonian splits as
\begin{equation}
H=\sum_{\mathbf{k}} H_{L,\mathbf{k}}+H_{R,\mathbf{k}}+H_{T,\mathbf{k}},
\end{equation} 
where
\begin{align}
H_{j=L/R,\mathbf{k}}=&\sum_{j\in L/R,\sigma} \big([-2\zeta\cos(k_y)-2\zeta\cos(k_z)] c_{j\sigma \mathbf{k}}^\dagger c_{j\sigma \mathbf{k}}\big) + \big(-\zeta c_{j+1\sigma \mathbf{k}}^\dagger c_{j\sigma \mathbf{k}}-\zeta c_{j\sigma \mathbf{k}}^\dagger c_{j+1\sigma \mathbf{k}}\big) \nonumber\\
&+ \big(\Delta_j(t)c^\dagger_{j\sigma \mathbf{k}}c^\dagger_{j\sigma' -\mathbf{k}} + \Delta_j^*(t)c_{j\sigma' -\mathbf{k}}c_{j\sigma \mathbf{k}}\big),
\end{align}
and
\begin{equation}
H_{T,k}= \sum_\sigma -\mathcal{T}\big(e^{i\frac{\phi(t)}{2}}c_{LN_{1,L}\sigma \mathbf{k}}^\dagger c_{R1\sigma \mathbf{k}} + e^{-i\frac{\phi(t)}{2}}c_{R1\sigma \mathbf{k}}^\dagger c_{LN_{1,L}\sigma \mathbf{k}}\big),
\end{equation}
with $k_y,k_z\in \{ -\pi,\pi \}$. The tunnel Hamiltonian preserves transverse wavevectors (specular tunneling). 

Subsequently, recalling that $\Delta$ depends only on $x$ ($\Delta_{x,\mathbf{k}}=N_k^2\delta_{k_y}\delta_{k_z}\Delta_{x}$), we self-consistently solve,
\begin{equation}
\begin{split}
G^<_{\mathbf{k}}=&G^r_{\mathbf{k}}\Sigma_{\mathbf{k}}^< G^a_{\mathbf{k}}\\
\Delta_{j_x}(t)=&\frac{1}{N_k^2}\Re \sum_{\mathbf{k}} igF^<_{j_x,j_x,\mathbf{k}}(t,t).
\end{split}
\end{equation}
Finally, the current is obtained using Eq. \eqref{If} for each subband, which are then summed up~\cite{sCuevas2006a}.

\subsection{Varying hopping/bandwidth}

\begin{figure*}[!htb]
\includegraphics[width=2.3in]{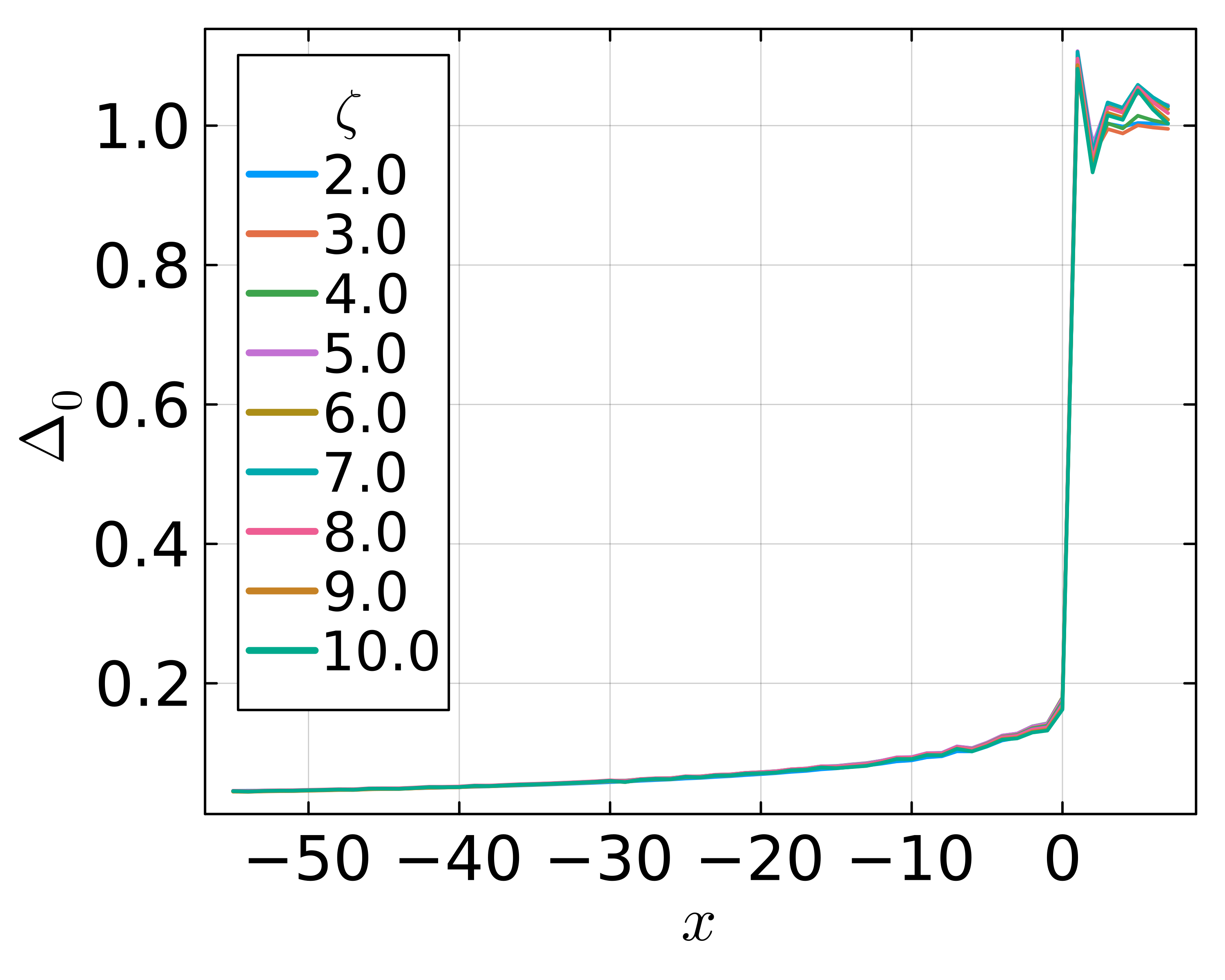}\llap{\parbox[b]{1.85in}{(a)\\\rule{0ex}{1.55in}}}\hspace*{1mm}
\includegraphics[width=2.3in]{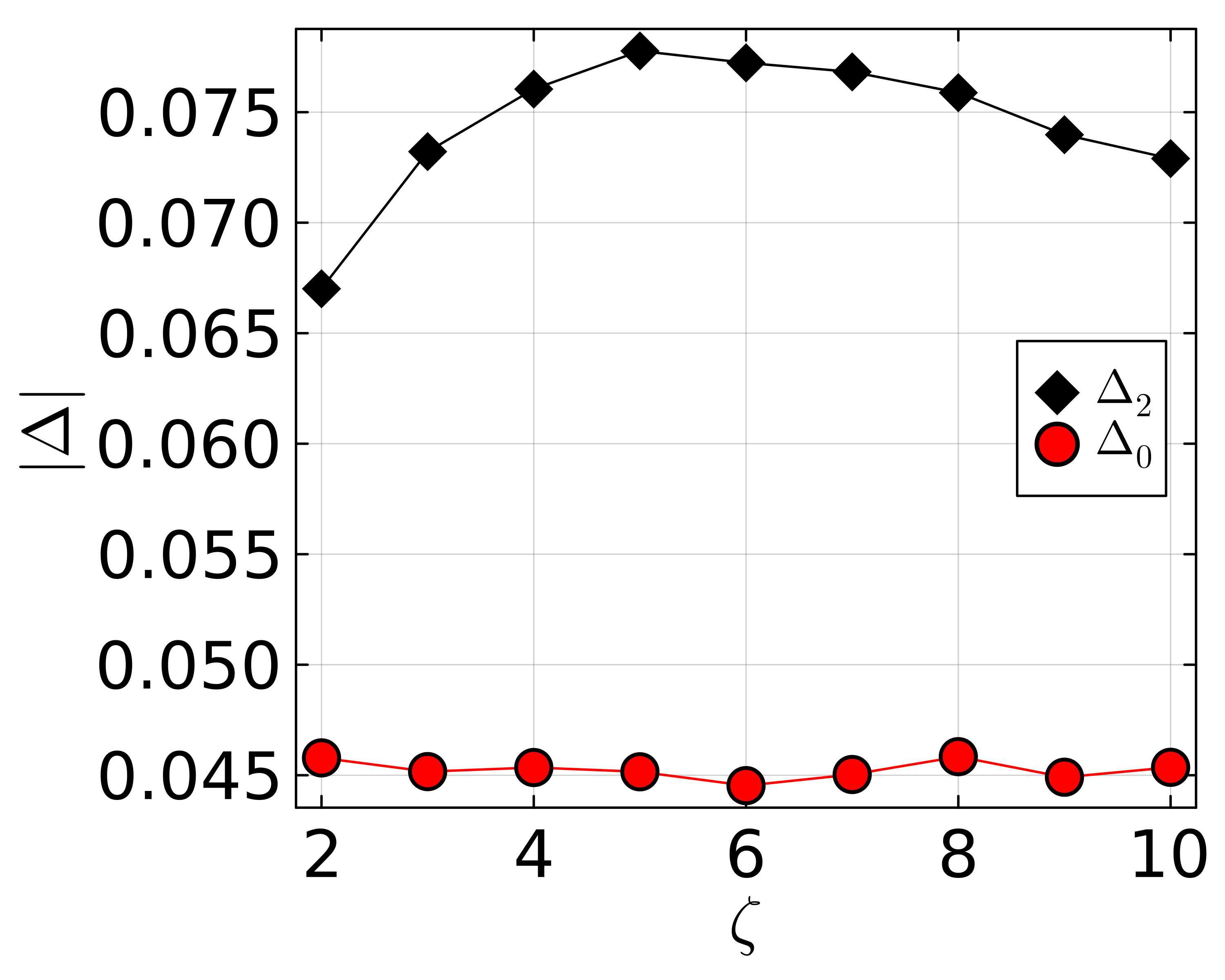}\llap{\parbox[b]{1.85in}{(b)\\\rule{0ex}{1.55in}}}\hspace*{1mm}
\includegraphics[width=2.3in]{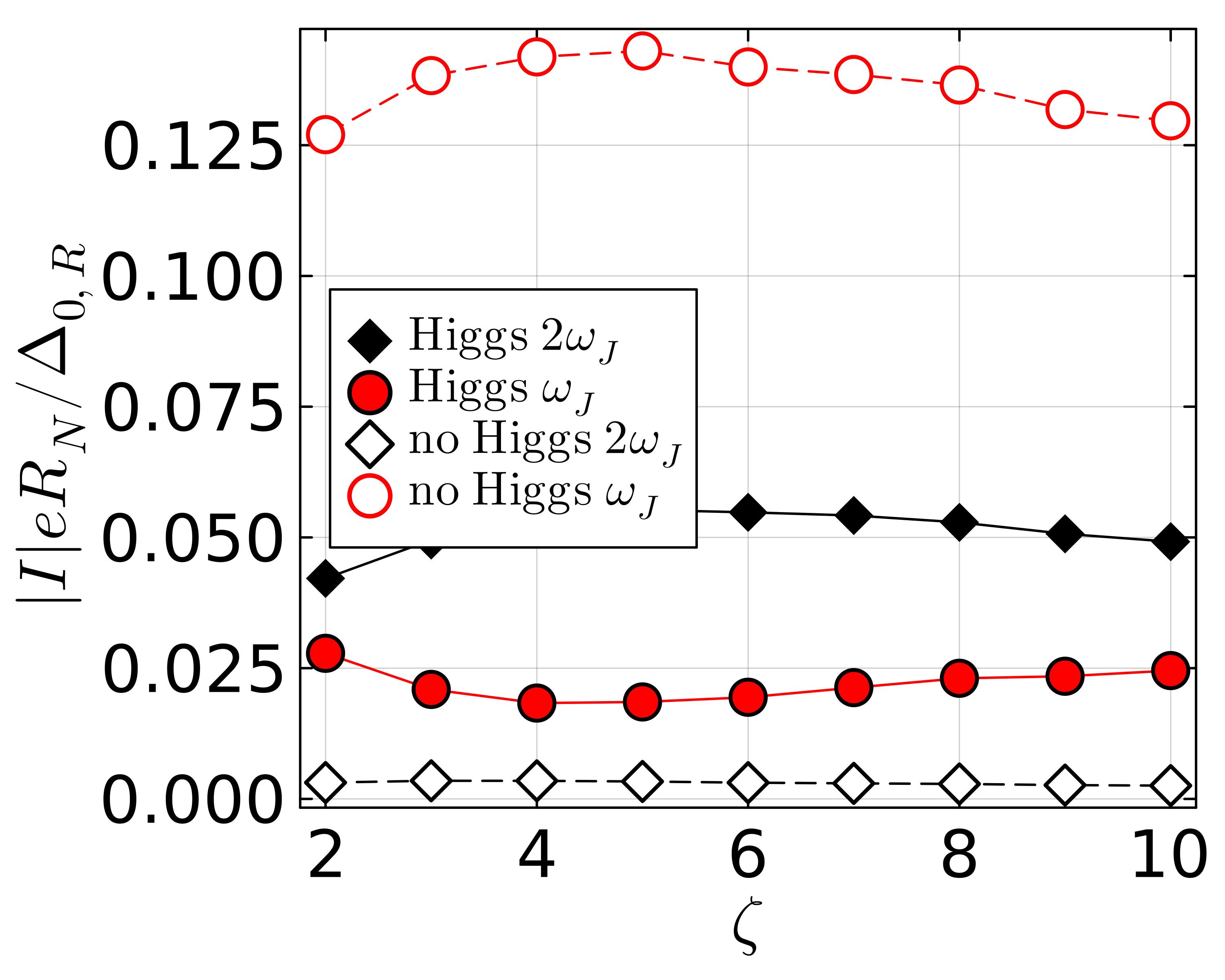}\llap{\parbox[b]{1.95in}{(c)\\\rule{0ex}{1.55in}}}
\caption{Numerically obtained results for varying $\zeta$ (bandwidth $12\zeta$), while keeping $\Delta_{0,L}=0.045$, $\Delta_{0,R}=1$, $eV=0.05$, and $\mathcal{T}/\zeta=0.4$ (normal state transparency$\ \approx 0.48$) fixed. We consider a three-dimensional junction with $N_L=56$, $N_R=7$, and $N_k=15$ channels in each of the two transverse directions. We use the three dimensional model as it allows us to explore much larger system lengths. This is needed as the coherence length is proportional to $\zeta$, necessitating larger system sizes. (a) The equilibrium gaps $\Delta_0$. We vary the BCS couplings with $\zeta$ suitably to keep the bulk values of $\Delta_0$ unchanged. (b) The non-equilibrium OP at frequency $\omega_J$, $|\Delta_2|$ at the first site on the left lead immediately neighbouring the junction, and the bulk equilibrium gap $\Delta_0$. (c) The amplitude (arbitrary units) of the $\omega_J$ (blue) and $2\omega_J$ (red) components of the current with (solid) and without (empty, dashed) Higgs oscillations. The Higgs-enhanced $2\omega_J$ current mirrors the trends of $\Delta_2$ in panel (b), remaining largely unaffected, whereas the usual $\omega_J$ current reduces as the DC value of the OP gets depleted.}
\label{FigS2} 
\end{figure*}

Here we present results with varying $\zeta$ (bandwidth = $12\zeta$), keeping the equilibrium gap $\Delta_{0,L/R}$ unchanged. We ensure that the junction transparency remains constant which amounts to fixing $\mathcal{T}/\zeta$. It is easy to understand this on considering the current flowing between two normal metallic leads~\cite{sCuevasbook}, which is proportional to the transparency $4( (\mathcal{T}/\zeta)^2/(1+(\mathcal{T}/\zeta)^2 )^2$. To keep the current unchanged with changing $\zeta$, one must keep $\mathcal{T}/\zeta$ constant. Physically speaking, the transparency constitutes the small parameter for the perturbative expansion of the current, and governs the low-energy transport properties. On scaling the tunnel coupling similarly as the hopping $\zeta$, electrons at the junction perceive the same barrier. Otherwise, we artificially end up in the tunnel limit with increasing bandwidth. 

With this in mind, we show that our results for the Higgs-mediated Josephson current remain unchanged with increasing bandwidth. This largely follows on noting that the non-equilibrium OP oscillating at frequency $\omega_J$ is essentially the result of the inverse proximity effect, albeit a dynamical one with time-dependent coupling to the other superconductor. As such, it is expected to depend on $\zeta$ only via the transparency, and on the two superconducting gaps. Indeed, in Fig.~\ref{FigS2}(b), we see that $|\Delta_2|$ is largely unaffected by varying $\zeta$, as we vary $\mathcal{T}$ and the BCS couplings suitably to keep the transparency and the bulk equilibrium gaps unchanged (Fig.~\ref{FigS2}(a)). The current operator introduces additional dependence on $\zeta$ via the tunneling self-energies, but once again, only through the transparency, which we keep unchanged. In the absence of the Higgs oscillations, to obtain this result, it is sufficient to look at the zero voltage Josephson harmonics~\cite{sBeenakker1991}. The coefficient of the first harmonic $\sin(\phi)$ yields $I_2\sim (\mathcal{T}/\zeta)^2$, while that of the second harmonic $\sin(2\phi)$ yields $I_4\sim (\mathcal{T}/\zeta)^4$. This dependence does not change with the bias voltage. In the presence of Higgs oscillations, two instances of the tunneling self-energy $\Sigma_T$ in the current operator at order $\mathcal{T}^4$ are replaced by the self energy $\Sigma_{\delta\Delta}\sim\delta\Delta$. As such, the bandwidth is not expected to play a major role, with the dominance of the Higgs-induced $2\omega_J$ current largely dependent on the transparency and equilibrium gap asymmetry. Note that there is no geometric dilution effect if we do not consider a point-contact or constriction, instead, we deal with a planar-junction geometry~\cite{sLee2023}.

\end{document}